\documentclass[aps,twocolumn,pra,superscriptaddress,longbibliography,nobalancelastpage]{revtex4-2}

\usepackage[english]{babel}
\usepackage{graphicx,nicefrac}
\usepackage{amssymb}
\usepackage{amsfonts}
\usepackage{xcolor}
\usepackage{amsmath}
\usepackage{siunitx}
\usepackage{soul}
\usepackage{comment}

\usepackage[utf8]{inputenc}

\usepackage{hyperref}

\usepackage{amsmath}
\usepackage{mathtools}
\usepackage{braket}
\usepackage{dsfont}
\usepackage{bigints}
\usepackage{graphicx}
\usepackage{amssymb}
\usepackage{xcolor}
\usepackage{bbold}

\usepackage{soul}

\graphicspath{ {./Images/} }

\hypersetup{
	colorlinks,
	linkcolor={red!95!black},
	citecolor={green!60!black},
	urlcolor={blue!95!black}
}

\newcommand{\approxpropto}{\mathrel{\vcenter{
			\offinterlineskip\halign{\hfil$##$\cr
				\propto\cr\noalign{\kern2pt}\sim\cr\noalign{\kern-2pt}}}}}

\newcommand{\nlchill}{NL-$\chi$LL }

\usepackage{adjustbox}
\makeatletter
\newcommand{\fitimage}[2][\@nil]{
	\begin{figure}
		\begin{adjustbox}{width=0.9\textwidth, totalheight=\textheight-2\baselineskip-2\baselineskip,keepaspectratio}
			\includegraphics{#2}
		\end{adjustbox}
		\def\tmp{#1}%
		\ifx\tmp\@nnil
		\else
		\caption{#1}
		\fi
	\end{figure}
}

\begin{document}

\title{Refermionized theory of the edge modes of a fractional quantum Hall cloud}
\author{Alberto Nardin}\email{alberto.nardin@unitn.it}
\author{Iacopo Carusotto}
\affiliation{Pitaevskii BEC Center, INO-CNR and Dipartimento di Fisica, Universit\`a di Trento, I-38123 Trento, Italy}

\begin{abstract}
Making use of refermionization techniques, we map the nonlinear chiral Luttinger liquid model of the edge modes of a spatially confined fractional quantum Hall cloud developed in our recent work [Phys. Rev. A {\bf 107}, 033320 (2023)] onto a one-dimensional system of massive and interacting chiral fermions, whose mass and interactions are set by the filling factor of the quantum Hall fluid and the shape of the external confining potential at the position of the edge.
As an example of the predictive power of the refermionized theory, we report a detailed study of the dynamic structure factor and of the spectral function of a fractional quantum Hall cloud. 
Among other features, our refermionized theory provides a physical understanding of the effective decay of the edge excitations and of the universal power-law exponents at the thresholds of the dynamic structure factor.
The quantitative accuracy of the refermionized theory is validated against a full two-dimensional calculation based on a combination of exact diagonalization and
Monte Carlo sampling.
\end{abstract}

\maketitle

\section{Introduction }\label{sec:introduction}
The fractional quantum Hall (FQH) effect is among the most fascinating concepts of modern quantum condensed matter physics~\cite{jainCompositeFermionsBook_2007,prangeQHE2012}, being a prototypical example of strongly correlated and incompressible quantum liquid exhibiting topological order~\cite{Wen_science_2019}.
Such a state of matter was originally observed more than $40$ years ago in semiconductor hetero-structures hosting a two-dimensional electron gas subject to a strong perpendicular magnetic fields~\cite{Tsui_PRL_1982}. Still now, it keeps attracting a strong attention due to its rich emergent physics, such as fractionally charged excitations~\cite{Laughlin_PRL_1983,Tsui_PRL_1982} with fractional~\cite{Arovas_PRL_1984} or even non-Abelian~\cite{Nayak_RMP_2008} exchange statistics, and gapless edge modes chirally propagating along the one-dimensional edge of the system~\cite{Wen_PRL_1990,Wen_PRB_1990b,Wen_PRB_1991, Wen_AdvPhys_1995,Wen_intJModPhysB_1992,Chang_RMP_2003}.
Over the years, these latter modes have been a powerful probe of the system, providing evidence of the topological order of the bulk~\cite{Wen_AdvPhys_1995} via the observation of fractional charge in shot-noise experiments~\cite{dePicciotto_nat_1997} and of non-trivial power-law exponents in the $I$-$V$ characteristic of tunneling junctions~\cite{Chang_PRL_1996, Chang_RMP_2003}, via thermal Hall conductance measurements~\cite{Banerjee_Nat_2018,Umansky_NatPhys_2023}, and, more recently, via measurements of the exchange statistics of the bulk quasiparticles~\cite{Nakamura_Nature_2020,Bartolomei_science_2020}. 
All these features are theoretically captured by a simple one-dimensional model, the chiral Luttinger liquid ($\chi$LL) theory~\cite{Wen_AdvPhys_1995}, which provides an accurate description of the edge dynamics in the long-wavelength, low-energy, weak excitation limit. 

While the FQH was originally observed in two-dimensional electron systems, a strong experimental attention is currently being devoted to the possibility of realizing FQH states in synthetic quantum matter systems, such as gases of ultra-cold atoms under synthetic magnetic fields~\cite{Cooper_2008,Bloch_RMP_2008,Cooper_RMP_2019,Goldman_RepProgPhys_2014} or fluids of strongly interacting photons in nonlinear topological photonics devices~\cite{Carusotto_NatPhys_2020,Carusotto_RMP_2013,Ozawa_RMP_2019}.
Important experimental steps in this direction have been recently reported in both atomic~\cite{Gemelke_2010,Fletcher_Science_2021,Tai_Nat_2017,Leonard_2022} and photonic systems~\cite{Schine_Nat_2016,Clark_Nat_2020,Roushan_NatPhys_2017}.
These setups are very interesting not only because they would allow for the realization
of novel FQH states of bosonic particles, but also because they typically offer more flexibility in the control of the system parameters as well as a wider variety of experimental probes as compared to the traditional transport and optical ones of electronic systems. In particular, while the  generation and diagnostics of neutral edge excitations in electronic systems requires ultrafast tools that are presently being developed with state-of-the-art electronic and optical technologies~\cite{Yusa_PRR_2022,Ashoori_PRB_1992,Frigerio_CommPhys_2024}, arbitrary space- and time-dependent potentials~\cite{Gauthier2016} can be straightforwardly applied to synthetic systems and high-resolution detection tools at the single-particle level are available~\cite{Gross2021}.

With an eye to these on-going experimental developments, we are carrying out a long-term project devoted to the study of the quantum dynamics of a FQH edge beyond the $\chi$LL description. As a first step, in~\cite{Nardin_EPL_2020} we investigated the linear and nonlinear dynamics of the edge of an Integer Quantum Hall (IQH) fluid. In the following work~\cite{Nardin_PRA_2023}, we extended the study to anharmonically trapped FQH clouds: based on the outcome of numerical calculations of the dynamics of a two-dimensional FQH fluid,
carried out via a novel combination of exact diagonalization with a Monte Carlo sampling of the matrix elements, 
we highlighted novel features such as a sizable group velocity dispersion and nonlinear effects for the edge modes. This allowed us to develop a generalized nonlinear chiral Luttinger liquid (NL-$\chi$LL) theory able to quantitatively reproduce the numerical predictions.

In this work, we make a further step forward in this long-haul program by showing how the \nlchill theory introduced in~\cite{Nardin_PRA_2023} can be mapped onto a model of one-dimensional interacting chiral fermions. This allows to use techniques originally developed for Tomonaga-Luttinger liquids to shine qualitative and quantitative light on crucial features of the edge dynamics such as the broadening of the dynamic structure factor (DSF) and the spectral function (SF) of the FQH fluid, the appearance of universal power law exponents at spectral edges, and the fine structure of the microscopic eigenstates. 



The structure of the article is the following.
In Sec.~\ref{sec:physical_system} we describe the physical system under investigation and we briefly review the \nlchill description we introduced in~\cite{Nardin_PRA_2023}. In Sec.~\ref{sec:refermionization} we present the reformulation of the problem in terms of a one-dimensional Tomonaga-Luttinger liquid model of massive and interacting chiral fermions. In Sec.\ref{sec:Applications} we make use of the refermionized model to shine physical light on the predictions of the \nlchill model and of the full 2D numerical calculations.
Conclusions are finally given in section \ref{sec:conclusions}.

Appendices contain a review of technical details of the methods and present additional numerical results. In particular, 
appendix \ref{appendix:MonteCarlo} reviews the numerical method used for the full two-dimensional simulations.
Appendix \ref{appendix:bosonization} reviews the main steps of the refermionization procedure used to write the nonlinear chiral Luttinger liquid theory in terms of a fermionic model. Appendix \ref{appendix:broadening} gives more information on the broadening of the density structure factor in a quartic trap. Appendix \ref{appendix:dsf_fitting} gives more details on the fitting technique used to extract the power-law exponents at the density structure factor thresholds. Appendix \ref{appendix:Jacks} presents additional results on the overlap of the \nlchill eigenstates with Jack polynomial wavefunctions. Appendix \ref{appendix:SpectralFunctionMC} describes the numerical Monte Carlo technique used to calculate the matrix elements of the spectral function.
Finally, appendix~\ref{appendix:sf} details the numerical calculation of these matrix elements within the \nlchill model.

\section{The physical system and the nonlinear chiral Luttinger liquid description}\label{sec:physical_system}
We consider a model of $N$ quantum particles of mass $m$ moving within a continuous two-dimensional plane, displaying short-range repulsive interactions, and subject to a uniform magnetic field $B$ orthogonal to the plane.
As usual, the single-particle states in a uniform $B$ organize in highly degenerate and uniformly separated Landau levels: in what follows, energies are measured in units of the cyclotron splitting between Landau levels $\hbar\omega_c=\hbar\, qB/m$ and lengths  in units of the magnetic length $\ell_B=\sqrt{\hbar/m\omega_c}$. 
Here $q$ is the charge of the particles. In the case of synthetic quantum matter, $qB$ is determined by the particular realization of the artificial gauge field. For example, in the case of rapidly rotating neutral atoms\cite{Cooper_2008}, $qB=2m\Omega_r$, $\Omega_r$ being the rotation frequency.
We also adopt the usual complex-valued shorthand $z=x+iy$.
    
Two-body interactions lift the macroscopic degeneracy of Landau levels and lead to the formation of highly-correlated incompressible ground states. 
The simplest examples of such states are the celebrated Laughlin states~\cite{Laughlin_PRL_1983,SimonWavefunctionology_2020}
\begin{equation}
    \label{eq_Laughlin}
    \Psi_L(\left\{z_i\right\})=\prod_{i<j}(z_i-z_j)^{1/\nu}\,\exp\left(-\frac{1}{4}\sum_{i}\left|z_i\right|^2\right),
\end{equation}
which entirely sit within the lowest Landau level. While the Laughlin state at filling $\nu=1/2$ is the exact ground state for contact-interacting bosons~\cite{WilkinGunn_PRL_1998}, other $\nu \neq 1/2$ Laughlin states are the exact ground state of certain bosonic or fermionic toy model Hamiltonians~\cite{Haldane_PRL_1983, TrugmanKivelson_PRB_1985, SimonRezayiCooper_PRB_2007_1,SimonRezayiCooper_PRB_2007_2} and turn out to be excellent approximations of the ground state of more realistic problems.
In the presence of a spatial in-plane confinement, the bulk of the FQH fluid maintains its incompressible nature, the ground state being separated by a finite energy gap from excited states, while the boundary hosts one-dimensional gapless chiral edge excitations~\cite{Wen_PRB_1991}.
    
In this work, we focus our attention on the edge excitations of a FQH fluid, with a particular attention to the $\nu=1/2$ Laughlin state which is most relevant for cold bosonic matter subject to a strong synthetic magnetic field.
These excitations correspond to chirally-propagating surface deformations of the incompressible cloud and, in the long-wavelength, low-energy, weak excitation limit, are accurately described by the $\chi$LL model~\cite{Wen_AdvPhys_1995, Wen_intJModPhysB_1992, Chang_RMP_2003, Cazalilla_PRA_2003}. 
On the other hand, the incompressible bulk only hosts gapped excitations, namely charged quasihole and quasielectrons~\cite{Laughlin_PRL_1983} and neutral magneto-roton-like density excitations~\cite{Girvin_PRB_1986}. The magnitude of the many-body (or Laughlin) energy gap separating them from the ground state is set by the interparticle interactions.

    \subsection{The nonlinear chiral Luttinger liquid description}
In our previous work~\cite{Nardin_PRA_2023} we considered a FQH cloud confined by a generic weakly anharmonic trap potential 
\begin{equation}
\label{eq:trap}
V_\text{conf}(r)=\lambda r^\delta.
\end{equation}
Under the assumption of a shallow trap and of moderate excitation strength, mixing with states above the many-body energy gap can be neglected and the dynamics is confined to the subspace of many-body wavefunctions obtained by multiplying the Laughlin wavefunction by holomorphic symmetric polynomials 
$P_{\kappa_l}\left(\left\{z_i\right\}\right)$ of the particle coordinates~\cite{Stone_PRB_1990,Wen_AdvPhys_1995,SimonWavefunctionology_2020},
    \begin{equation}
        \label{eq:laughlinEdge}
        \Psi_{\kappa_l}(\left\{z_i\right\}) = P_{\kappa_l}\left(\left\{z_i\right\}\right)\,\Psi_L(\left\{z_i\right\});
\end{equation}
{here the index $\kappa_l$ labels a partition of the integer $l$. See appendix~\ref{appendix:MonteCarlo} for additional details.}
As long as we consider polynomials $P_{\kappa_l}$ of degree $l$ much smaller than the number of particles, these wavefunctions {provide a complete basis for the description of the} excitations of the edge of the FQH cloud { carrying $l$ additional units of angular momentum with respect to the Laughlin ground state Eq.~\eqref{eq_Laughlin}}.

The temporal dynamics of the cloud was simulated by restricting to the class of states \eqref{eq:laughlinEdge}, evaluating the matrix elements by Monte Carlo sampling of the wavefunctions, and then performing the time-evolution on the restricted Hamiltonian with standard numerical techniques. A brief review of the method is reported in App.~\ref{appendix:MonteCarlo}. 
The result of these full 2D numerical calculations was used to inspect the linear and non-linear dynamics of the edge modes in response to a spatially-dependent and pulsed excitation. 

This allowed us to identify new features of the dynamics beyond the standard $\chi$LL theory and, in particular, we showed that the low-energy, long-wavelength, weak-excitation physics of the FQH edge is successfully captured by a \nlchill Hamiltonian of the form
\begin{equation}
        \label{eq:NLXLL_hamiltonian}
    	\hat{H}=\int d\theta\left(\pi\,\frac{\Omega}{\nu}\,\hat{\rho}^2+\frac{c}{\nu^2}\,\frac{2\pi^2}{3}\,\hat{\rho}^3-\pi\frac{\beta_\nu}{\nu}\frac{c}{R_{cl}}\,\left(\partial_\theta\hat{\rho}\right)^2\right)
\end{equation}
where normal ordering of the operator products is assumed, even if not explicitly written, and the edge-density operators $\hat{\rho}$ obey the standard bosonic commutation relations~\cite{Wen_AdvPhys_1995}
\begin{equation}
        \label{eq:KacMoodyCommutator}
        [\hat{\rho}(\theta),\hat{\rho}(\theta')]=-i \frac{\nu}{2\pi} \partial_\theta\delta(\theta-\theta')
\end{equation}
of a $\chi$LL. 
In addition to { the standard $\chi$LL term $\propto\hat\rho^2$}, the Hamiltonian Eq.~\eqref{eq:NLXLL_hamiltonian} includes { new terms, which are irrelevant in the renormalization group sense  in the low-energy limit but are crucial for the correct description of the system's dynamics at fixed and large enough momenta: a} boson-boson interactions proportional to $\hat{\rho}^3$ and a surface-tension-like term proportional to $(\partial_\theta\hat{\rho})^2$, which causes a modification to the dispersion of linear waves. { The effect of these terms on the dynamics of realistic finite-size systems will be the focus of this work.}
The parameter $\beta_\nu$ is a filling-dependent constant characteristic of Laughlin FQH states. From the microscopic calculations we performed it turns out to be approximately equal to 
\begin{equation}
    \beta_\nu\simeq\frac{\pi}{8}\frac{1-\nu}{\nu}.
\end{equation} 
While no solid theoretical explanation is yet available for this remarkably simple form, we conjecture that it may be justified by relating it to the viscosity of quantum Hall fluids~\cite{Avron_PRL_1995} and the gapped magneto-roton excitations of the bulk~\cite{Girvin_PRB_1986}.
    
The quantitative agreement of the \nlchill theory with the full 2D numerical calculation is all the way more surprising given that it is a straightforward generalization of Wen's $\chi$LL theory that, for a given topological state of filling factor $\nu$, only involves two non-universal parameters $\Omega$ and $c$ characterizing the confinement potential at the position of the cloud's edge. These parameters physically correspond to the angular velocity of the edge modes 
\begin{equation}
    \label{eq:angular_velocity}\Omega=R_{cl}^{-1} \partial_r V_\text{conf}|_{R_{cl}}=\lambda \,\delta\, R_{cl}^{\delta-2},
\end{equation}
i.e. the force exerted by the confinement potential [Eq.~\eqref{eq:trap}] at the classical radius $R_{cl}=\sqrt{2N/\nu}$ of the cloud, and to the radial gradient 
\begin{equation}
    \label{eq:curvature}
    c=R_{cl}^{-1} \partial_r \Omega|_{R_{cl}}=\lambda\, \delta (\delta-2)\, R_{cl}^{\delta-4}
\end{equation} 
of this force.
In this work, we will restrict our attention to the case of positive-curvature confinement potentials with $c>0$. 
The different behaviours that occur for other forms of the potential will be explored in a forthcoming work.
            
In view of following developments, it is useful to note how the evolution of the edge-density operator under the \nlchill Hamiltonian is the quantum analog of a Korteweg-de Vries equation of classical fluid dynamics
    \begin{equation}
        \label{eq:qKdV}
    	\partial_t \hat{\rho} = 
    	i[\hat{H},\hat{\rho}]=
    	-\Omega \partial_{\theta}\hat{\rho}
    	-\frac{\pi c}{\nu}\,\partial_{\theta}\hat{\rho}^2
    	-\alpha\partial_{\theta}^3\hat{\rho},
    \end{equation}
    {where for notational convenience we introduced the shorthand 
    \begin{equation}
        \alpha=\beta_\nu {c}/{R_{cl}}.
    \end{equation}
    }
    For the sake of completeness, it is worth noting that non-universal corrections to the $\chi$LL Hamiltonian analogous to Eq.\eqref{eq:NLXLL_hamiltonian} have been investigated also in the different context of reconstructed edges~\cite{KunYang_PRL_2003}. Here, however, the non-trivial dispersion was the result of a gradient expansion of the non-local Coulomb interaction between electrons rather than an intrinsic property of strongly correlated FQH liquids with short-range interactions. Finally, a summary of all possible irrelevant perturbations to the $\chi$LL Hamiltonian allowed by symmetries was extensively discussed  from a conformal-field theory perspective in~\cite{Fern_PRB_2018}.

	{ 
	\subsection{Thermodynamic limit} 
	We here briefly comment how the system is properly scaled up so as to reach a well defined thermodynamic limit of $N\gg 1$. 
	By introducing a one-dimensional particle density $\hat\sigma = \hat\rho/R_{\rm cl}$ and position $\xi=R_{\rm cl}\theta$, we can rewrite the commutator as 
    \begin{equation}
    [\hat{\sigma}(\xi),\hat{\sigma}(\xi')]=-i \frac{\nu}{2\pi} \partial_\xi\delta(\xi-\xi'),
\end{equation}
    and the Hamiltonian Eq.~\eqref{eq:NLXLL_hamiltonian} as
	\begin{equation}
		\hat{H}\!=\!\!\int \!\! d\xi \left(\!\pi\,\frac{\Omega R_{\rm cl}}{\nu}\,\hat{\sigma}^2+c R_{\rm cl}^2\left(\frac{2\pi^2}{3\nu^2}\,\hat{\sigma}^3-\pi\frac{\beta_\nu}{\nu}\left(\partial_\xi\hat{\sigma}\right)^2\right)\right);
	\end{equation}
	it is clear	that the model will be $N$ independent as long as $\Omega R_{\rm cl}$ and $c R_{\rm cl}^2$ are kept fixed while $N$ is varied: what matters therefore are the confinement force setting the propagation speed $\Omega R_{\rm cl}$ along the edge and its gradient $c R_{\rm cl}^2$ at the system's edge.	
	It is clear that fixing both is not possible when a potential in the form of a single monomial $V_{\rm conf}=\lambda r^\delta$ is confining a circular FQH cloud; more general potentials $V_{\rm conf}(r)$ lifting this restriction can however be devised with no further difficulty.
	
	While the speed $\Omega R_{\rm cl}$ only contributes an overall rigid slope to the energy dispersion, the scaled curvature parameter $c R_{\rm cl}^2$ is what matters for the non-linear physics that is of interest here. In the following, in our numerical simulations performed in the disk geometry, this is the parameter we are going to keep fixed when increasing the system's size.
	}

\section{The refermionization procedure}\label{sec:refermionization}

Starting from the \nlchill theory reviewed in the previous section, we now show how the quantum dynamics of the edge excitations of a FQH cloud can be exactly mapped onto an equivalent model of one-dimensional, massive and interacting chiral fermions, for which 
one can make use of the artillery of techniques developed in the context of non-linear Luttinger liquids~\cite{ImambekovGlazman_Science_2009, ImambekovGlazman_RMP_2012}.

{For numerical convenience, we will focus on a disk geometry, and therefore work with the angular density $\hat\rho$.}
By rescaling the bosonic field as $\hat\rho'=\hat\rho / \sqrt{\nu}$, the Hamiltonian of Eq.~\eqref{eq:NLXLL_hamiltonian} can be written as
\begin{equation}
    \label{eq:NLXLL_hamiltonian_rescaled}
	\hat{H}=\int d\theta\left(\pi\,\Omega\,\hat{\rho}'^2+\frac{c}{\sqrt{\nu}}\,\frac{2\pi^2}{3}\,\hat{\rho}'^3-\pi\alpha\left(\partial_\theta\hat{\rho}'\right)^2\right),
\end{equation}
where the commutation relations Eq.~\eqref{eq:KacMoodyCommutator} are correspondingly rescaled to
\begin{equation}
    \label{eq:KacMoodyCommutator_rescaled}
    [\hat{\rho}'(\theta),\hat{\rho}'(\theta')]=-\frac{i}{2\pi} \partial_\theta\delta(\theta-\theta'),
\end{equation}
which is the standard commutation rule of bosonized density modes in a Tomonaga-Luttinger model~\cite{Haldane_JPC_1981, Giamarchi_QP1D_2004}. 

The similarity with the Tomonaga-Luttinger model does not stop here, and standard bosonization identities~\cite{Giamarchi_QP1D_2004} can be used to show~\footnote{Note that the refermionization map is not complete as $\psi = e^{i\phi/\nu}\rightarrow e^{i\phi'/\sqrt{\nu}}$ is not a single fermion annihilation operator. In spite of this, the above mapping works as long as we compute density-related observables.} that the full rescaled Hamiltonian Eq.~\eqref{eq:NLXLL_hamiltonian_rescaled} is the bosonized version of a model of one-dimensional (chiral) fermions. 
{ The technical details of this refermionization procedure are summarized in Appendix~\ref{appendix:bosonization}.}
This leads to a fermionic Hamiltonian of the form
\begin{equation}
    \label{eq:fermion_hamiltonian}
	\hat{H}'=\sum_l \epsilon_l \hat R^\dagger_l \hat R^{\vphantom{\dagger}}_l - \frac{\alpha}{2}\sum_l l^2 \hat{\rho}_{-l}{\!\!\!\!\!'\,\,\,\,}\hat{\rho}_{l}{\!'},
\end{equation}
where the free-fermion dispersion
\begin{equation}
    \label{eq:fermion_diseprsion}
    \epsilon_l = \Omega l + \frac{l(l-1)}{2m^*}
\end{equation}
has a quadratic contribution with an effective mass $m^*=(c/\sqrt{\nu})^{-1}$ and interactions occur via the short-range potential $V_{12}(\theta_1-\theta_2)=2\pi\,\alpha\, \delta''(\theta_1-\theta_2)$. As usual, the fermionic creation (annihilation) operators $\hat R_l^\dagger$ ($\hat R_l$) obey anticommutation rules
\begin{equation}
    \{R^{\vphantom{\dagger}}_l,R^\dagger_{l'}\}=\delta_{l,l'}\,,
\end{equation}
and $\hat \rho_l{\!'}={\int e^{-i l \theta }\hat\rho'd\theta =}\sum_{l'} \hat R^\dagger_{l'-l} \hat R^{\vphantom{\dagger}}_{l'}$ is the Fourier transform of the density operator $\hat\rho{'=\hat R^\dagger(\theta)\hat R^{\vphantom{\dagger}}(\theta)}$ { and $\hat R^{\vphantom{\dagger}}(\theta) = \frac{1}{\sqrt{2\pi}}\sum_{l} e^{i l \theta}\hat R_l^{\vphantom{\dagger}}$}.

It is interesting to note that the term proportional to $\hat\rho^3$ describing the interactions between the bosonic modes in Eq.~\eqref{eq:NLXLL_hamiltonian} translates into the non-interacting mass term in the refermionized Eq.~\eqref{eq:fermion_diseprsion}. 
Vice-versa, the quadratic term proportional to $(\partial_\theta\hat\rho)^2$ describing the group velocity dispersion of the bosons translates into an interaction term in the fermionic picture. 

In the regime we are investigating here the ground state $\ket{0}$ of the fermionic Hamiltonian Eq.~\eqref{eq:fermion_hamiltonian} is a Fermi sea filling all the states below the Fermi point $l_F=0$
\begin{equation}
\label{eq:FermiSea}
\begin{cases}
R_{l\leq0}^\dagger\ket{0}=0\\
R_{l>0}\ket{0}=0
\end{cases}
\end{equation} 
and is the only available state for its value of the angular momentum~\footnote{Notice that higher-order terms in the fermion free-particle dispersion are in principle required to make this ground-state well defined, which we however neglect mainly for two reasons: first of all, these corrections are expected not to be relevant at the energy scales we are considering; secondly, we have not characterized these higher order terms numerically~\cite{Nardin_PRA_2023}.}. 
{ Of course the interaction term in Eq.~\eqref{eq:fermion_hamiltonian} is to be intended as normally ordered with respect to this Fermi sea ground state.} 

\begin{figure}[htbp]
   	\begin{adjustbox}{width=.5\textwidth, totalheight=\textheight-2\baselineskip,keepaspectratio,right}
      	\includegraphics[]{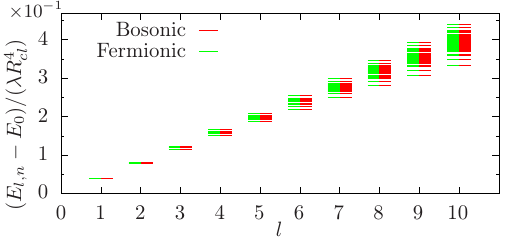}
    \end{adjustbox}
    \vspace{0.0cm}\caption{{Numerical results for the energy spectrum $E_{l,n}-E_0$ of the bosonic \nlchill model Eq.~\eqref{eq:NLXLL_hamiltonian} (green lines) and its fermionic counterpart Eq.~\eqref{eq:fermion_hamiltonian} (red lines).    
    Eigenstates are plotted against their angular momentum $l$. 
    We here used $N=25$ bosons at $\nu=1/2$ filling in a quartic trap; the model free parameters, $\Omega$ and $c$ have been calculated from $N$ and $\nu$ using Eq.~\eqref{eq:angular_velocity} and Eq.~\eqref{eq:curvature}.
    }}\label{fig:comparison}
\end{figure}

{
In Fig.~\ref{fig:comparison} we numerically check that the refermionized model Eq.~\eqref{eq:fermion_hamiltonian} is as expected (up to numerical precision) isospectral with the bosonic \nlchill Eq.~\eqref{eq:NLXLL_hamiltonian}.
}

To conclude this Section, it is important to note that density-related observables of the physical FQH system directly map onto the density operator of the refermionized model, showing that the charge-zero sector of the FQH edge theory maps (at low energies) onto a chiral generalization of the non-linear Luttinger model of one-dimensional fermions~\cite{ImambekovGlazman_Science_2009,ImambekovGlazman_RMP_2012}.
On the other hand, the creation/annihilation operators of the physical particles forming the FQH fluid cannot be mapped in a simple way to fermionic creation/annihilation operators. 
As we are going to see in the next Section, this mathematical fact makes the refermionization approach 
a useful tool for characterizing { theoretically} the DSF -- and similarly all density-related observables -- much more than the SF.




\section{Applications}
\label{sec:Applications}
After having introduced in the previous Section the general formalism of the refermionization procedure, we are now going to apply it to the study of two physical quantities of actual experimental interest, in particular the DSF characterizing the spatio-temporal dynamics of edge excitations and, then, the SF characterizing the energetics of particle removal processes.

\subsection{Dynamic structure factor}
The linear response of the edge-density to a perturbation that couples to the edge-density operator is encoded in the DSF for the edge modes, mathematically defined as
\begin{equation}
    \label{eq:dynamic_structure_factor}
    S_l(\omega)=\int \frac{dt}{2\pi}\,e^{i\omega t}\braket{e^{i\hat{H}t} \delta\hat{\rho}_{l} e^{-i\hat{H}t} \delta\hat{\rho}_{-l}}\,.
\end{equation}
Here, $\delta\hat\rho_l$ is the $l$-th component of the angular Fourier transform of the edge-density variation with respect to the Laughlin ground state, defined as
\begin{equation}
    \delta\hat{\rho}(\theta) = \int_0^\infty (\hat{\rho}-\braket{\hat{\rho}}_{GS})\,r\,dr.
\end{equation}

\begin{figure}[htbp]
   	\begin{adjustbox}{width=.5\textwidth, totalheight=\textheight-2\baselineskip,keepaspectratio,right}
      	\includegraphics[]{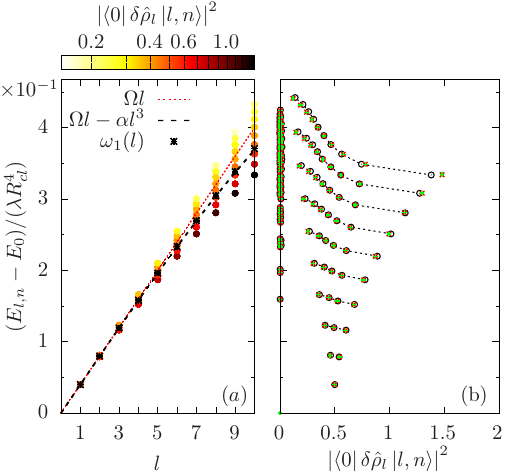}
    \end{adjustbox}
    \vspace{0.0cm}\caption{(a) Numerical results for the energy spectrum $E_{l,n}-E_0$ of a FQH cloud confined by a quartic $V_{\rm conf}(r)=\lambda r^4$ trap. Eigenstates are plotted against their angular momentum $l$ and are coloured according to their DSF weight $|\bra{0}\delta\hat{\rho}_l\ket{l,n}|^2$.
    The red dotted line is the linear dispersion of the edge modes, $\Omega l$; the black line includes the cubic correction, $\Omega l-\alpha l^3$. Black circles with crosses indicate the center-of-mass \eqref{eq:DSFFirstMoment} of the DSF.
    (b) Plot of the DSF (on the $x$-axis) against the excitation energies $E_{l,n}-E_0$ (on the $y$-axis). 
    The black circles are the result of full 2D numerical calculations, the red crosses are the predictions of the { bosonic} \nlchill~model of Eq.~\eqref{eq:NLXLL_hamiltonian}
    { and the green plus symbols show the result of the exact diagonalization of the refermionized description Eq.~\eqref{eq:fermion_hamiltonian}.}
    For each angular momentum $l$, the main emerging structures have been joined with black-dashed lines as a guide for the eye.
    Both panels are for $N=25$ bosons at $\nu=1/2$ filling.
    \label{fig:Energy_DSF}}
\end{figure}

As long as we neglect the coupling to states above the many-body energy gap and we restrict to excitation frequencies $\omega$ below the many-body energy gap, the DSF~\eqref{eq:dynamic_structure_factor} can be rewritten by introducing a projector onto the relevant low-energy subspace, which gives
\begin{equation}
    \label{eq_laughlin_projected_DSF}
    S_l(\omega)=\sum_n\,\delta\left(\omega-\omega_{l,n}\right) \left|\bra{0}
    \delta\hat{\rho}_l \ket{l,n}\right|^2\,.
\end{equation}
Here, $n$ runs through the excited states of angular momentum $l$ and $\omega_{l,n}=E_{l,n}-E_0$ is their energy difference from the Laughlin ground state.
In our previous work~\cite{Nardin_PRA_2023}, we computed this quantity both via a full two-dimensional numerical calculation of the FQH cloud { through Monte Carlo techniques analogues to those detailed in App.~\ref{appendix:MonteCarlo}}, and via { exact-diagonalization of} the { full bosonic one-dimensional} \nlchill~model of Eq.~\eqref{eq:NLXLL_hamiltonian}. 
As it is displayed in Fig.~\ref{fig:Energy_DSF}, the results of the two methods are in excellent quantitative agreement. 
{ Since the refermionization procedure is exact, the numerical prediction obtained from exact-diagonalization of the full fermionic model Eq.~\eqref{eq:fermion_hamiltonian} (green plus symbols) perfectly recover those of the bosonic theory (red crosses).}

We here wish to highlight the importance of the dispersion and interaction terms in Eq.~\eqref{eq:NLXLL_hamiltonian}.
Even though for a fixed $l$ most of the DSF weight is concentrated in the lowest energy state, a non-zero weight is present in higher energy states as well. As a result, the dispersion of linear waves, defined as the center-of-mass of the DSF
\begin{equation}
    \label{eq:DSFFirstMoment}
     \omega_1(l)=\frac{\int \omega S_l(\omega) d\omega}{\int S_l(\omega) d\omega},
 \end{equation}
does not coincide with the lowest energy state at $E_{-}(l)$.
It is henceforth not possible to identify this state with a single boson excitation at the same angular momentum $l$, as previously done~\cite{Stone_PRB_1992,Wan_PRB_2003,Wan_PRB_2008,Jolad_PRL_2009,Jolad_PRB_2010}.
The boson-boson interaction term proportional to $\hat\rho^3$  in the \nlchill~model of Eq.~\eqref{eq:NLXLL_hamiltonian} is therefore playing a crucial role in spreading the DSF over a finite range of energies.
Furthermore, as we are going to see in the next Sections, the fact that $\omega_1(l)$ deviates from a linear behaviour
highlights the importance of the dispersive term proportional to $(\partial_\theta \hat\rho)^2$.




\begin{figure}[htbp]
   	\begin{adjustbox}{width=.5\textwidth, totalheight=\textheight-2\baselineskip,keepaspectratio,right}
      	\includegraphics[]{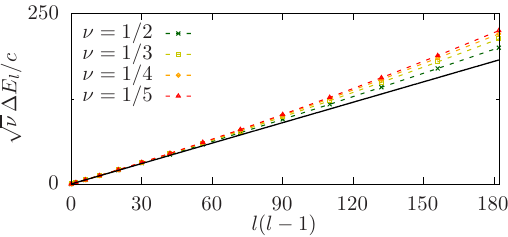}
    \end{adjustbox}
    \vspace{0.0cm}\caption{{ \nlchill numerical results for the DSF broadening $\Delta E_l$ (points connected by dashes), as a function of $l(l-1)$ - $l$ being the angular momentum. 
    The numerically computed broadening $\Delta E_l$ is compared to its free fermion approximation (solid black line), Eq.~\eqref{eq:broadening1}, for different values of the filling fraction $\nu$. 
    Analogously to Fig.~\ref{fig:comparison}, the model free parameters $\Omega$ and $c$ have been calculated from $N=25$ and $\nu$ using Eq.~\eqref{eq:angular_velocity} and Eq.~\eqref{eq:curvature}.}}
    \label{fig:broadening}
\end{figure}

\subsubsection{Broadening of the dynamic structure factor}\label{subsec:dsf_broadening}

As a first application of the refermionized Hamiltonian Eq.~\eqref{eq:fermion_hamiltonian}, we look at the broadening of the DSF that is well visible in Fig.~\ref{fig:Energy_DSF} and manifests as a progressive spreading of the energies with the angular momentum $l$ of the cloud.

In order to get a simple picture of the underlying physics, we start by making the approximation of neglecting the interactions between fermions described by the last term of the Hamiltonian~\eqref{eq:fermion_hamiltonian}. 
{ This approximation is well justified at low-energies/long-wavelengths, since the $\hat \rho^3$ term in Eq.~\eqref{eq:NLXLL_hamiltonian}, from a renormalization group perspective, has a smaller scaling dimension than the $(\partial_\theta\hat\rho)^2$ one~\cite{Dubail_PRB_2012,Fern_PRB_2018}.}
This leaves us with a free fermion model of dispersion $\epsilon_l$ whose excitations { $\ket{l,n}$} consist of particle-hole pairs around the Fermi level, taken to be at $l_F=0$. 
Within this approximation, for a given value of the angular momentum $l$ of the excitation, the DSF { of Eq.~\eqref{eq_laughlin_projected_DSF} gets its contributions from single particle-hole excitations, since $\hat\rho_l$ annihilates a single particle-hole pair. 
At angular momentum $l$, there are $l$ such states: namely,  $\ket{l,n}=\hat R^\dagger_{n+l} \hat R^{\vphantom{\dagger}}_{n}\ket{0}$ for $-l < n \leq 0$. 
In the free-fermion limit, on these single particle-hole states, $\left|\bra{0}\delta\hat{\rho}_l \ket{l,n}\right|^2=\nu$ and thus the DSF}
has a flat profile { (not shown here)} in between 
{ the states with the smallest and largest excitation energies, which (due to the concavity of $\epsilon_l$) are $\ket{-}=\hat R^\dagger_{1} \hat R^{\vphantom{\dagger}}_{-l+1}\ket{0}$ and $\ket{+}=\hat R^\dagger_{l} \hat R^{\vphantom{\dagger}}_{0}\ket{0}$ respectively. The DSF can thus be explicitly written as
\begin{equation}
    \label{eq:free_fermion_DSF}
    S_l(\omega) = \nu \sum_{n=-l+1}^0 \delta\left(\omega - (\epsilon_{l+n}-\epsilon_n)\right).
\end{equation}
}
{The excitation} energies {associated to the two thresholds} are
\begin{eqnarray}
\label{eq:energies}
E_-(l)&=&\epsilon_1-\epsilon_{1-l} \\
E_+(l)&=&\epsilon_{l}-\epsilon_0,
\end{eqnarray}
{
and the normalized first moment of the DSF $\omega_1(l)$ Eq.~\eqref{eq:DSFFirstMoment}, within this approximation, coincides with the average
\begin{equation}
    \label{eq:DSFcenter}
	\omega_1(l)=\frac{E_++E_-}{2}=\Omega l.
\end{equation}
 This result roughly agrees with the numerically computed  averaged dispersion, shown in Fig.~\ref{fig:Energy_DSF}(a); the small deviation with respect to this free-fermion approximation signals } the importance of the fermion interaction term, or, equivalently, of the additional bosonic dispersive term{. Such a fermionic interaction term is indeed responsible for the cubic frequency shift highlighted in Fig.~\ref{fig:Energy_DSF}(a) and for the markedly non-flat and asymmetric DSF profile visible in Fig.~\ref{fig:Energy_DSF}(b). We will further comment on this in the following}.

{ In spite of this deviation on the DSF first moment, the non-interacting fermion approximation is able to well-capture the broadening of the DSF, as we are now going to see. An estimate for this quantity can be obtained by}
taking the difference of the energies in Eq.~\eqref{eq:energies}
\begin{equation}
    \label{eq:broadening1}
    \Delta E_l = E_+(l)-E_-(l)=\frac{c}{\sqrt{\nu}}\,l(l-1),
\end{equation}
which is proportional to the trap curvature parameter $c$ and grows quadratically with the excitation angular momentum $l$. Interestingly, the prefactor depends on the filling fraction of the underlying FQH state.
A more quantitative measure of the DSF broadening is provided by the second moment of the DSF 
\begin{equation}
\label{eq:omega2}
    \omega_2(l) = \frac{\int (\omega - \omega_1(l))^2 S_l(\omega)\, d\omega}{\int S_l(\omega) d\omega}\,.
\end{equation}
{ In the non-interacting fermions regime, using Eq.~\eqref{eq:free_fermion_DSF} and Eq.~\eqref{eq:DSFcenter}, this easily evaluates to}
\begin{equation}
    \label{eq:broadening2}
    \omega_2(l) = \frac{c^2}{\nu}\,\frac{l^2(l^2-1)}{12}.
\end{equation}

{
The approximated broadening measure Eq.~\eqref{eq:broadening1} is successfully compared with the numerical \nlchill calculation in Fig.~\ref{fig:broadening}; it can be seen that the free-fermion dispersion correctly captures the broadening at small angular momenta $l$, specially for the largest fillings, since the ratio of the fermion mass term to fermion-fermion interactions roughly scales $\propto(1-\nu)^{-1}$ (see Eq.~\eqref{eq:fermion_hamiltonian}).}

\begin{figure}[htbp]
   	\begin{adjustbox}{width=.5\textwidth, totalheight=\textheight-2\baselineskip,keepaspectratio,right}
      	\includegraphics[]{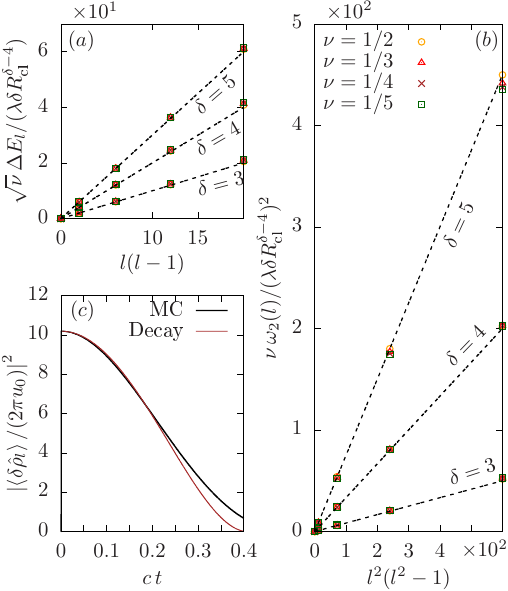}
    \end{adjustbox}
    \vspace{0.0cm}\caption{(a) Plot of the (normalized) total width $E_+(l)-E_-(l)$ defined in Eq.~\eqref{eq:broadening1} as a function of the (rescaled) angular momentum $l(l-1)$. 
    The results of full 2D numerical calculations for various values of $\nu$ and $\delta$ accurately follow the behaviour $\sqrt{\nu}\Delta E_l/(\lambda \delta R_{cl}^{\delta-4})=(\delta-2)l(l-1)$ predicted by Eq.~\eqref{eq:broadening1} together with Eq.~\eqref{eq:curvature}, shown here as black-dashed lines.
    (b) Plot of the second moment $\omega_2(l)$ of the DSF as a function of the (rescaled) angular momentum $l^2(l^2-1)$. 
    The results of the full 2D numerical calculations for various values of $\nu$ and $\delta$ accurately follow the behaviour $\nu\omega_2(l) / (\lambda\delta R_{cl}^{\delta-4})^2 = (\delta-2)^2l^2(l^2-1)/12$ predicted by Eq.~\eqref{eq:broadening2} together with Eq.~\eqref{eq:curvature}, shown here as black-dashed lines.
    (c) Time-evolution of the square modulus of the fundamental component of the edge-density response in response of a weak pulsed excitation of strength $u_0$ and angular momentum $l=3$. 
    The black curve is the result of the full 2D numerical calculation, while the brown curve is the the short-time decay prediction of Eq.~\eqref{eq:density_decay}. In all panels, a FQH cloud of $N=25$ particles is considered.    
    \label{fig:broadening_and_decay}}
\end{figure}

{ After having compared the approximated expression Eq.~\eqref{eq:broadening1} with numerical \nlchill results, we compare} the two broadening measures Eq.~\eqref{eq:broadening1} and Eq.~\eqref{eq:broadening2} with {microscopic Monte Carlo data for the full 2D problem,} for different values of the exponent $\delta$ of the confinement potential and of the filling factor $\nu$
in Fig.~\ref{fig:broadening_and_decay}(a) and  Fig.~\ref{fig:broadening_and_decay}(b) respectively.
Also in this case, the analytical predictions appear to quite accurately capture the numerical data { at the relatively small values of the imparted angular momentum $l$ here analyzed}, especially for { large} values of the filling fraction $\nu$, and of the confinement exponent $\delta$: this regime allows in fact to minimize effects caused by the gradient of the curvature parameter $c$, which is neglected in our \nlchill~theory. 
As a further evidence in support of our conclusions, more numerical results on the quartic $\delta=4$ case are shown in Appendix \ref{appendix:broadening}.


It is interesting to highlight that the formulas Eq.~\eqref{eq:broadening1} and Eq.~\eqref{eq:broadening2} for the DSF broadening, obtained neglecting interactions between fermions, remain 
accurate well outside the regime where interactions are irrelevant. 
The non-interacting fermion approximation predicts in fact the DSF to be flat and centered at the linear $\chi$LL dispersion [Eq.~\eqref{eq:DSFcenter}] $\omega_1(l)=\Omega l$; however, as one can see in Fig.~\ref{fig:Energy_DSF}, while $\omega_1(l)$ only slightly deviates from the linear behaviour over the values of $l$ considered in Fig.~\ref{fig:broadening_and_decay}, the DSF rapidly starts
acquiring a highly non-trivial lineshape much before the non-interacting fermion approximation for the broadening breaks down.
Analogously, while the broadening $\Delta E_l$ in Eq.\eqref{eq:broadening1} compares remarkably well with the full 2D numerical calculation, the threshold energies $E_\pm$ predicted by the non interacting model Eq.~\eqref{eq:energies} separately do not.

\subsubsection{Spectroscopic remarks}
From an experimental perspective, besides the spectroscopic measurements of~\cite{Fabbri_PRA_2015} the second moment Eq.~\eqref{eq:broadening2} of the DSF can be indirectly measured by looking at the short- and moderate-time part of the temporal decay of edge-density excitations on top of the FQH cloud.
Consider that the cloud is excited via a time-dependent perturbation $U(\theta,t)$ whose strength is almost constant in the vicinity of the classical radius. 
Using linear perturbation theory, 
we obtain the following result for the edge-density response 
\begin{equation}
    \label{eq:perturbative_edge_response}
\braket{\delta\hat{\rho}(\theta,t)}=\frac{1}{\pi}\Im\left[
    \sum_l e^{i l \theta}
    \int \widetilde{U}_{l}(\omega) \,S_l(\omega) \,e^{-i\omega t}\,d\omega \right]
\end{equation}
where $\widetilde{U}_{l}(\omega)$ is the Fourier transform of the excitation potential and $S_l(\omega)$ the DSF defined in Eq.~\eqref{eq:dynamic_structure_factor}.

Assuming that the spectrum $\widetilde{U}_{l}(\omega)$ of the perturbation is approximately constant across the peak of $S_l(\omega)$ (this requires that the excitation pulse is sufficiently short compared to the characteristic time-scale of the edge dynamics), we can approximate the integral appearing on the right-hand side of \eqref{eq:perturbative_edge_response} as
\begin{equation}
    \widetilde{U}_{l}(\omega_1(l)) e^{-i\omega_1(l) t} \int \,S_l(\omega) \,e^{-i(\omega-\omega_1(l)) t}\,d\omega.
\end{equation}
Up to not-too-large times, the exponential inside this integral can be expanded to second order, which gives
\begin{multline}
     \widetilde{U}_{l}(\omega_1(l)) e^{-i\omega_1(l) t} \times \\ \times \int \,S_l(\omega) \,\left(1 - \frac{(\omega-\omega_1(l))^2 t^2}{2}\right)\,d\omega\,.
\end{multline}
Then, using the sum-rule $\int S_l(\omega)\,d\omega=\nu l\,\Theta(l)$ for the static structure factor of a FQH cloud in the thermodynamic limit ($\Theta$ is here the Heaviside step function), we finally get
\begin{multline}
 \label{eq:density_decay}
\braket{\delta\hat{\rho}(\theta,t)}\simeq -\frac{\nu}{\pi}\sum_{l>0}\left(1 - \frac{\omega_2(l) t^2}{2}\right)\times\\ \times\frac{\partial}{\partial\theta} \,\Re \left[
     e^{i (l \theta-\omega_1(l) t)}
    \widetilde{U}_{l}(\omega_1(l))\right].
\end{multline}
At short times after the excitation, the decay of density modes follows a quadratic law. Its time-scale is set by the second moment of the DSF which, in turn, depends on the curvature parameter $c$ and on the filling fraction of the bulk, $\nu$ according to Eq.\eqref{eq:broadening2}. 
As a sidenote, notice that it is indeed $\omega_1(l)$ [Eq.~\eqref{eq:DSFFirstMoment}] which sets the propagation velocity of each $l$ mode, including the group velocity dispersion effect due to the dispersive contribution.

The accuracy of the approximated expression Eq.~\eqref{eq:density_decay} at short times is successfully validated against the exact evolution in Fig.~\ref{fig:broadening_and_decay}(c), where we plot the response of the edge modes to an excitation carrying a well-defined angular momentum. These results support 
a physical interpretation of the edge excitation decay as the result of the decoherence of the different particle-hole excitations that we originally proposed in~\cite{Nardin_EPL_2020} for the IQH case.

\begin{figure}[t]
   	\begin{adjustbox}{width=.45\textwidth, totalheight=\textheight-2\baselineskip,keepaspectratio,right}
      	\includegraphics[]{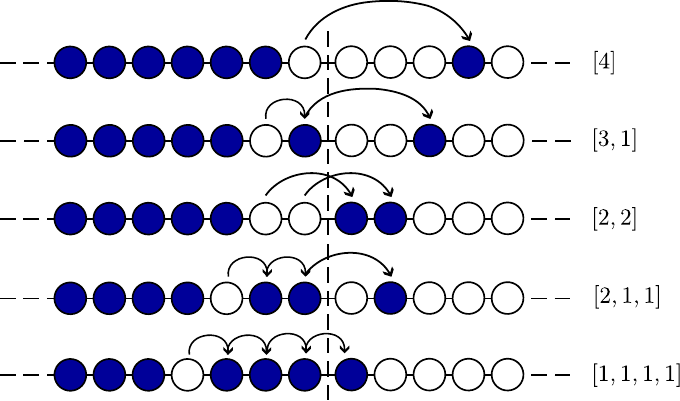}
    \end{adjustbox}
    \vspace{0.0cm}\caption{
    Schematic diagram of the different particle-hole excitations (black-arrows) across the Fermi point (vertical-dashed lines) at a fixed angular momentum $l=4$. For each state, the corresponding partition of $l=4$ are indicated on the right-hand side of the plot.
    \label{fig:ParticleHole}}
\end{figure}

\subsubsection{Threshold singularities of the dynamic structure factor}\label{subsec:dsf_exponent}

On top of the broadening effect discussed in the previous Subsection, the curves for the DSF plotted in Fig.~\ref{fig:Energy_DSF}(b) for growing $l$ { markedly deviate from the flat DSF associated to free-fermions that we just discussed, and }
clearly show the appearance of some peculiar singular behaviours 
close to the spectral thresholds. { These features emerge when interactions are not disregarded from the fermionic model Eq.~\eqref{eq:fermion_hamiltonian}, and can qualitatively be understood due to the fact that the previously discussed bare particle-hole excitations can interact with the Fermi sea in non-trivial ways, leading to the non-flat DSF numerically observed in Fig.~\ref{fig:Energy_DSF}(b).}
Such features are { indeed} reminiscent of those emerging from the theory of non-linear (non-chiral) Luttinger liquids~\cite{ImambekovGlazman_RMP_2012}.

The main challenge in the theoretical study of both the bosonic and fermionic models of  Eq.~\eqref{eq:NLXLL_hamiltonian} and Eq.~\eqref{eq:fermion_hamiltonian} comes from the fact that both the group velocity dispersion and the non-linearity are proportional to the same curvature parameter $c$. For this reason, perturbative approaches based on a hydrodynamic formulation where dispersion dominates over interactions~\cite{PriceLamacraft_PRB_2014, PriceLamacraft_arxiv_2015} do not give consistent results. On the other hand, many features of the fermionic theory Eq.~\eqref{eq:fermion_hamiltonian}, in particular the behaviour around the energy thresholds, can be be successfully studied making use of so-called ``mobile-impurity" approaches. 

\begin{figure}[htbp]
   	\begin{adjustbox}{width=.5\textwidth, totalheight=\textheight-2\baselineskip,keepaspectratio,right}
      	\includegraphics[]{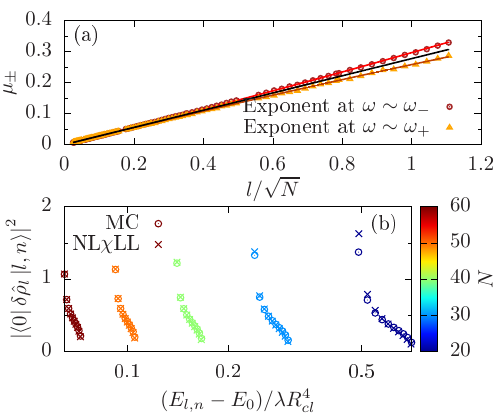}
    \end{adjustbox}
    \vspace{0.0cm}\caption{(a) Plot of the DSF threshold exponents $\mu_\pm$ as a function of the excitation momentum $q=l/R_{cl}\propto l/\sqrt{N}$, numerically extracted from the predictions of the \nlchill model of Eq.~\eqref{eq:NLXLL_hamiltonian} by means of a power-law fit to the DSF data close to the lower and upper threshold. The
    extracted exponents at the lower and upper thresholds are shown as red circles and yellow triangles, respectively, and are quite successfully compared to the mobile-impurity result of Eq.~\eqref{eq:exponentDSF} shown as a black line. An even better agreement is found including a phenomenological quadratic correction as done in Eq.~\eqref{eq:ExponentDeviation} (red and brown lines).
    (b) Comparison between the microscopic DSF weigths extracted from full 2D numerical calculations (circles)
    and those obtained from the \nlchill model Eq.~\eqref{eq:NLXLL_hamiltonian} (crosses) for different particle numbers $N$ (colorscale) at a fixed angular momentum $l=10$. The DSF weights are plotted against the excitation energies in a logarithmic scale.
    \label{fig:DSFexponents}}
\end{figure}

Within a non-interacting fermion model, the lower threshold corresponds to { the state $\ket{-}=\hat R^\dagger_{1} \hat R^{\vphantom{\dagger}}_{-l+1}\ket{0}$, i.e. to } the displacement of a particle from deep below the Fermi point to right above it. 
Correspondingly, the upper threshold corresponds to 
{ the state $\ket{+}=\hat R^\dagger_{l} \hat R^{\vphantom{\dagger}}_{0}\ket{0}$, i.e. to } the displacement of a particle initially located at the Fermi point up to high energy states. 
These processes can be seen as the creation of a deep-hole or a high-energy-particle accompanied by a slight shift of the Fermi point (sketched in the last or first row of Fig.~\ref{fig:ParticleHole}, respectively).  
Once we include again interactions, if we focus on the threshold regions, the interacting fermion theory of Eq.~\eqref{eq:fermion_hamiltonian} can be replaced by an effective two-band model, namely a (chiral) Luttinger liquid at $l\sim0$ and a single deep-hole/high-energy-particle at $l$~\cite{Pustilnik_PRL_2006,Khodas_PRB_2007, ImambekovGlazman_RMP_2012}. This latter then acts as an impurity off which particles close to the Fermi point, that is the Luttinger liquid, can perform small-momentum-transfer scattering processes. As it is discussed in detail in~\cite{Pustilnik_PRL_2006}, 
such processes lead to a power-law enhancement of the DSF close to the lower-energy threshold $\omega_-(l)$ and a corresponding power-law suppression at the high-energy threshold $\omega_+(l)$.
In formulas, we have that
\begin{equation}
	\label{eq:dsf_threshold_behavior}
    \begin{cases}
        S_l(\omega\sim\omega_-) \propto \theta(\omega-\omega_-)\left(\frac{1}{\omega-\omega_-}\right)^{\mu_-}\\
        S_l(\omega\sim\omega_+) \propto \theta(\omega_+-\omega)\left(\omega_+-\omega\right)^{\mu_+}
    \end{cases}
\end{equation}
where the exponents
\begin{equation}
    \label{eq:exponentDSF}
    \mu_+\simeq\mu_-\simeq \frac{2\alpha l}{c/\sqrt{\nu}} = 2\beta_\nu\sqrt{\nu}\,\frac{l}{R_{cl}}
\end{equation}
only depend on the excitation momentum $l/R_{cl}$ and the bulk filling fraction $\nu$, { since $\beta_\nu\simeq \frac{\pi}{8}\frac{1-\nu}{\nu}$ only depends on the filling fraction~\cite{Nardin_PRA_2023}. In particular, they do not depend on} the specific values of the non-universal trap parameters $\Omega$ and $c$. 
Even though a finite value of $c$ is essential for the emergence of the singular power-law behavior, the values of the exponents { (modulo its momentum dependence)} turn out to be universal properties of strongly correlated FQH fluids with {\it short-range} interactions, a manifestation of strong bulk correlations extending all the way through the edge. 
This peculiar universal behavior emerges because the exponent is set by $2m^*\alpha$, but both the effective mass $m^*$ and the interaction strength proportional to $\alpha$ emerge because of the presence of the trap causing a gradient of angular velocity at the cloud's edge. 
So, while $\alpha$ is directly proportional to $c$, the effective mass $m^*$ is inversely proportional to it: the curvature parameter $c$ therefore cancels out from the exponents $\mu_\pm$.
{We however stress here that this ``universality" crucially relies on the fact that interactions among the FQH constituents are short-range. We expect it to be spoiled in a more general case - e.g. for an electron gas, where Coulomb interactions cannot be ignored.}


A direct validation of the power-law behaviour via full two-dimensional simulations of the FQH clould is made difficult by the fast, almost exponential ($\log\mathcal{D}\propto \sqrt{l}$) growth of the Hilbert space dimension $\mathcal{D}$ at the large angular momentum $l$ values that are needed to interpolate power-law behaviours at both thresholds.

In our previous work~\cite{Nardin_PRA_2023} and in the previous Sections, we have seen a quantitative agreement between full two-dimensional numerical simulations and the predictions of the \nlchill theory of Eq.~\eqref{eq:NLXLL_hamiltonian}. This statement is further validated in Fig.~\ref{fig:DSFexponents}(b) where we compare the DSF weights for increasing number of particles $N$ at fixed angular momentum $l$: as the excitation wavevector $q=l/R_{cl}\propto l/\sqrt{N}$ decreases the \nlchill model becomes even more accurate, so we can expect it to correctly account for the behaviour of the DSF in the thermodynamic limit. On this basis, we restrict our numerical analysis to the \nlchill theory which grants us access to the large $l$ values that are needed to precisely extract the power law exponents { numerically from the DSF data. The fitting procedure is detailed in Appendix ~\ref{appendix:dsf_fitting}}.

The numerical predictions for the exponents at the lower {(red circles)} and higher thresholds {(yellow triangles)} are plotted in Fig.~\ref{fig:DSFexponents}(a): a good agreement with the analytical prediction of Eq.~\eqref{eq:exponentDSF} {(black line)} is found for small values of the wavevector $q$, with quadratic corrections at higher $q$ phenomenologically compatible with the form 
\begin{equation}
    \label{eq:ExponentDeviation}
    \mu_\mp=\pm\left[\left(1\pm \frac{\alpha l}{c/\sqrt{\nu}} \right)^2-1\right],
\end{equation}
{ shown as red/brown lines in Fig.~\ref{fig:DSFexponents}(a).}

These results illustrate the power of the refermionized theory in capturing the peculiar behaviour of the edge excitations of a FQH cloud 
and show that the FQH edges indeed behave as a peculiar example of nonlinear Luttinger liquid~\cite{ImambekovGlazman_RMP_2012}.

\subsubsection{Fine structure of the dynamic structure factor}\label{subsec:DSF_fine_structure}

\begin{figure}[htbp]
   	\begin{adjustbox}{width=.5\textwidth, totalheight=\textheight-2\baselineskip,keepaspectratio,right}
      	\includegraphics[]{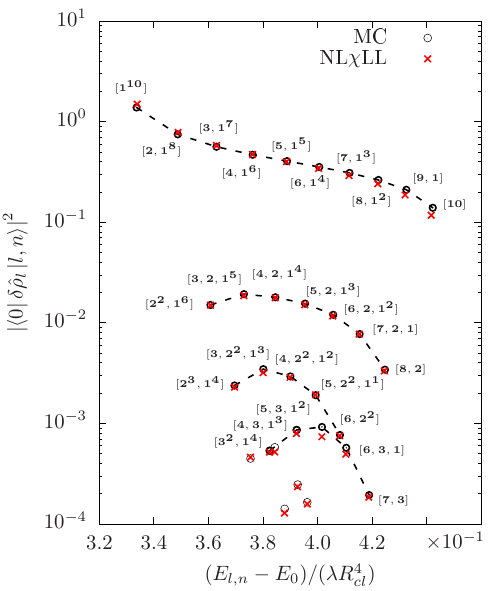}
    \end{adjustbox}
    \vspace{0.0cm}\caption{Plot of the DSF weights $|\bra{0}\delta\hat\rho_l\ket{l,n}|^2$ against the excitation energy $E_{l,n}-E_0$ for a $l=10$ excitation of a FQH cloud of $N=25$ bosons at  filling $\nu=1/2$ confined by a quartic $\delta=4$ trap. The black circles are the result of a full 2D numerical calculation and are compared to the predictions of the \nlchill~model of Eq.~\eqref{eq:NLXLL_hamiltonian}. Differently from Fig.~\ref{fig:Energy_DSF}(b), the weights are plotted here in logarithmic scale.
    All points with a sizable DSF weight are labeled in terms of the partition $\eta$ corresponding to the particle-hole excitation which has a strongest overlap with the eigenstate.
    \label{fig:DSFLogScale}}
\end{figure}

Going beyond the mesoscopic quantities investigated in the previous Subsections, it is interesting to see how the refermionized theory provides an interesting physical interpretation also for the microscopic structure of the eigenstates, in particular the values of the matrix elements $|\bra{0}\delta\hat\rho_l\ket{l,n}|^2$ entering the DSF. As an example, a plot of these matrix elements is displayed in Fig.~\ref{fig:DSFLogScale} for the $l=10$ case.

Within the fermionic model, we expect that at the level of the free fermion approximation, the $\delta\hat\rho_l$ operator only connects the ground state Fermi sea to the single particle-hole states on top of it (first, second, fourth and fifth rows of Fig.~\ref{fig:ParticleHole}) and that the $l$ such transitions have the same amplitude.
This simple picture is of course modified by the presence of fermionic interactions, so that non-zero matrix elements may appear also for states corresponding to several particle-hole excitations (third line of Fig.~\ref{fig:ParticleHole}). In spite of these corrections, the overall validity of the free fermion picture is fully confirmed by the numerical data shown in Fig.~\ref{fig:DSFLogScale}:  
the interacting eigenstates are found to maintain a dominant weight on the particle-hole basis of non-interacting fermions. On this basis, in Fig.~\ref{fig:DSFLogScale} we keep labelling the states in terms of the partition $\eta$ of the non-interacting particle-hole state which has the largest weight on the eigenstate. As usual, a partition $\eta=[\eta_1,\eta_2,\eta_3\ldots]$ is defined to have $\eta_1\geq\eta_2\geq\eta_3\geq\ldots$ and corresponds to a state where, starting from a filled Fermi sea, the highest energy particle is promoted by $\eta_1$ orbitals, the second-highest one by $\eta_2$ orbitals and so on: for the sake of clarity, a few examples of partitions and of the corresponding states are illustrated in Fig.~\ref{fig:ParticleHole}.  Further evidence in support of the free fermion picture is displayed in Fig.\ref{fig:FermionOverlaps} of Appendix \ref{appendix:Jacks}.

From Fig.~\ref{fig:DSFLogScale}, it is apparent how the matrix elements 
tend to organize in a hierarchical way. Besides the principal sequence of highest-weight states corresponding to single particle-hole excitations discussed above, 
well-distinguishable secondary sequences of states are visible, carrying a much weaker DSF weight. For all these data-points, an excellent agreement is found between the \nlchill model and the full two-dimensional calculation. 

\begin{figure}[htbp]
   	\begin{adjustbox}{width=.5\textwidth, totalheight=\textheight-2\baselineskip,keepaspectratio,right}
      	\includegraphics[]{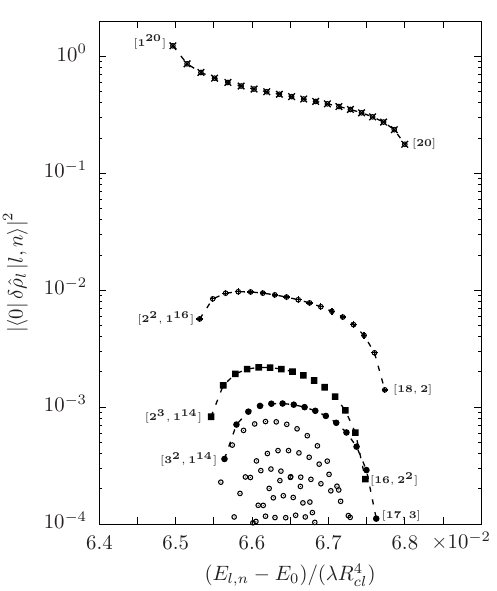}
    \end{adjustbox}
    \vspace{0.0cm}\caption{
    Plot of the DSF weights $|\bra{0}\delta\hat\rho_l\ket{l,n}|^2$ against the excitation energy $E_{l,n}-E_0$ for a $l=20$ excitation of a FQH cloud of $N=300$ bosons at filling $\nu=1/2$ confined by a quartic $\delta=4$ trap. The points have been computed via the \nlchill~model of Eq.~\eqref{eq:NLXLL_hamiltonian}.
    \label{fig:ExtendedDSFLogScale}}
\end{figure}

While this latter method gets quickly impracticable for larger values of $l$, the \nlchill model provides a manageable and reliable approach up to much larger $l$ values. An example of such calculation is displayed in Fig.\ref{fig:ExtendedDSFLogScale}: thanks to the larger $l=20$ value, a larger number of secondary structures is clearly discernible.
Interestingly, these structures can be grouped together (black-dashed lines) by considering squeezing processes in which the high-energy fermion looses one unit of angular momentum by exciting one more fermion across the Fermi point that delimits the Fermi sea of filled states, e.g.  $[7,3]\rightarrow[6,3,1]\rightarrow[5,3,1,1]\rightarrow\hdots[3,3,1,1,1,1]$.

As a final point, it is interesting to look at these structure from the point of view of the two-dimensional FQH cloud. Remarkably, we find a large overlap ($\gtrsim 95\%$ for the considered system sizes) 
between the eigenstates of the 2D FQH system and
Jack polynomial states~\cite{Macaluso_PRA_2017,Macaluso_PRA_2018} labeled by the \textit{same} partitions $\eta$. 
The interested reader can find a color plot of the overlap matrix in Fig.\ref{fig:quarticTrapJacksOverlaps} of Appendix~\ref{appendix:Jacks}, together with a brief description of the numerical method used to compute them. 
While no complete explanation of this remarkable result is available yet, it hints at a deep relation between the Pauli principle for the fermions of the refermionized theory and the generalized Pauli principle naturally implemented by the Jacks~\cite{BernevigHaldane_PRL_2008}. 
A complete understanding of this relation will be the subject of future work.

\subsection{Spectral function }\label{subsec:spectral_function}

Another quantity of great interest in the study of Luttinger liquids is the spectral function (SF), which describes the probability of removing a particle from the system at a given energy.
This quantity is relevant for the study of particle tunneling into $\chi$LL and is connected to power-laws in the current-voltage characteristic of FQH systems~\cite{Wen_PRB_1991b}. Recently it was shown to contain relevant information on Haldane fractional exclusion statistics~\cite{CooperSimon_PRL_2015} and to be directly accessible to experiments in various synthetic setups~\cite{CooperSimon_PRL_2015,Umucalilar_PRA_2017}.

Focusing on the bosonic Laughlin state at $\nu=1/2$, in this Subsection we show how the results of the full two-dimensional calculation of the SF are quantitatively captured by \nlchill model of Eq.~\eqref{eq:NLXLL_hamiltonian} once this is supplemented with the bosonized form of the particle annihilation operator~\cite{Wen_intJModPhysB_1992}
\begin{equation}
    \hat{\psi}(\theta) = e^{-i\hat{\phi}(\theta)/\nu}
    \label{eq:bosonized_annihil}
\end{equation}
in terms of a bosonic phase operator $\hat{\phi}$ related to the density through $\hat{\rho}=-\partial_\theta\hat{\phi}/2\pi$~\cite{Wen_intJModPhysB_1992, Wen_AdvPhys_1995}. The refermionized approach is then used to shine physical light on the peculiar properties of the SF.

\subsubsection{Comparison with full 2D numerical calculations}
As usual, we define the SF as
\begin{equation}
    \label{eq:spectralFunction}
    A_l(\omega)=\sum_{f} \left| \bra{f}\hat{a}_{(N-1)/\nu-l}\ket{0}\right|^2\delta(\omega - \omega_{f,0}).
\end{equation}
where the sum runs over all $N-1$ particle states $\ket{f}$ and $\ket{0}$ is the $N$ particles Laughlin ground state. Here, $\hat{a}_{(N-1)/\nu-l}$ annihilates a particle with angular momentum $(N-1)/\nu-l=\Delta l$. The reason behind this notation for the angular momentum will be clarified shortly.

Let us first notice that, provided the anharmonic confinement does not induce mixing of the low-lying edge excitations with states above the many-body gap, the frequency-integrated SF $A_l=\int d\omega\,A_l(\omega)$ at fixed $l$ is independent of the specific confinement as the eigenstates $\ket{f}$ are connected by a unitary transformation. The standard $\chi$LL result $A_l\propto l^{1/\nu-1}$ (at small values of $l$) is therefore maintained in spite of the SF being broadened and having a highly non-trivial line-shape. As such, we focus here on the energy-resolved $A_l(\omega)$.

The initial $N$ particle Laughlin state has total angular momentum $L_0^{(N)}=N(N-1)/(2\nu)$. The occupied single-particle orbital of largest angular momentum has angular momentum $(N-1)/\nu$, so the angular momentum $L_f$ of the final state after removing one particle lies in the range
\begin{equation}
\frac{(N-1)(N-2)}{2\nu} \leq L_f \leq \frac{N(N-1)}{2\nu}\,,
\end{equation}
that is, between $L_0^{(N-1)}$ and $L_0^{(N)}$.
For convenience, we will focus our attention on the removal of a particle close to the edge of the system. Since such a particle carries 
a large angular momentum $\lesssim (N-1)/\nu$, the final state can be seen as a low-angular-momentum edge-excitation with angular momentum $l$ with respect to the $N-1$ particles-Laughlin ground state; therefore, the relevant matrix elements can be studied through the \nlchill~model of \eqref{eq:NLXLL_hamiltonian}.

\begin{figure*}[htbp]
   	\begin{adjustbox}{width=1\textwidth, totalheight=\textheight-2\baselineskip,keepaspectratio,center}
      	\includegraphics[]{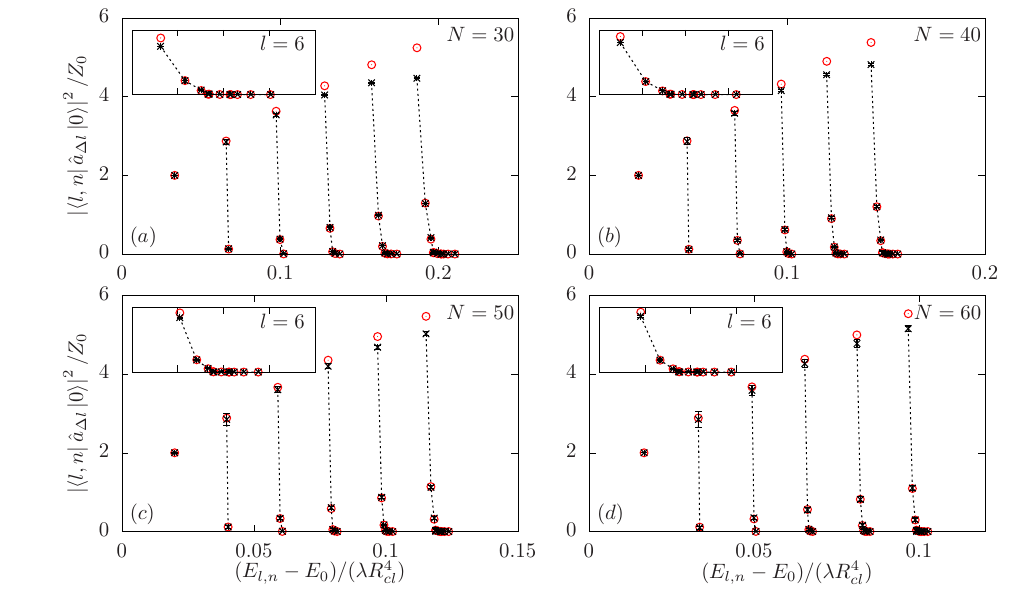}
    \end{adjustbox}
    \vspace{0.0cm}\caption{SF weight $|\bra{f}\hat a_{(N-1)/\nu-l}\ket{0}|^2$, normalized to the $l=0$ Laughlin-Laughlin matrix element $Z_0$. Black points are the result of the full 2D numerical calculation for bosons at a constant filling $\nu=1/2$ in a quartic $\delta=4$ trap but different numbers of particles in each panel: (a) $N=30$, (b) $N=40$, (c) $N=50$ and (d) $N=60$. The red circles display the predictions of the \nlchill model of Eq.~\eqref{eq:NLXLL_hamiltonian}.
    The insets display a magnified view of the $l=6$ curves.
    The amplitudes relative to each  $l$-sector ($l=1\hdots6$) have been joined with black dashed lines as a guide for the eye.
    The point corresponding to $l=0$ is not shown as it equals $1$ by definition due to the normalization factor.
    \label{fig:spectralFunction}}
\end{figure*}

Within our theoretical framework, we calculate the matrix elements appearing in Eq.~\ref{eq:spectralFunction} through a Monte Carlo sampling of the full two-dimensional wavefunctions, as described in Appendix~\ref{appendix:SpectralFunctionMC}.
The results of this microscopic calculation are shown in Fig.~\ref{fig:spectralFunction} and are compared to the { numerical} prediction of the \nlchill model{, obtained by diagonalizing numerically the \nlchill Hamiltonian Eq.~\eqref{eq:NLXLL_hamiltonian}}. In this latter calculation, the single-particle destruction operators $\hat{a}_{\Delta l}$ of the full two-dimensional theory are interpreted within the $\chi$LL framework as
\begin{equation}
	\label{eq:annihilation_operator}
\hat{a}_{\Delta l} = \int d\theta \,\frac{e^{i\, \Delta l\, \theta}}{\sqrt{2\pi}} \,\hat{\psi}(\theta)\,,    
\end{equation}
where the bosonized form Eq.\eqref{eq:bosonized_annihil}  of the annihilation operator is used.
In order to remove an overall normalization factor which ambiguously depends on the cutoff length-scale of the effective edge-boson theory~\cite{PalaciosMacDonald_PRL_1996,Jolad_PRB_2007,Jolad_PRL_2009,Jolad_PRB_2010,Giamarchi_QP1D_2004}, all matrix elements have been normalized to the Laughlin-Laughlin transition matrix element $Z_0 = |\bra{0'}a_{(N-1)/\nu}\ket{0}|^2$, {where $\ket{0'}$ is the $N-1$ particle ground-state. Additional details on the numerical calculation of the SF via diagonalization of the \nlchill model can be found in Appendix~\ref{appendix:sf}.} 

As expected, only a limited number of states within each angular momentum $L_f=L_{0}^{(N-1)}+l$ sector have a significant matrix element.
A good qualitative and quantitative agreement between the two theories is clearly visible in Fig.~\ref{fig:spectralFunction}, in particular for the low-energy states at low $l$. At large $l$, a good agreement is recovered as the system is made larger and the wavelength gets correspondingly longer. Most interestingly, these numerical results prove the correctness of the exponential expression Eq.\eqref{eq:bosonized_annihil} for the particle annihilation operator in terms of the bosonic field within our \nlchill model Eq.~\eqref{eq:NLXLL_hamiltonian}.

\subsubsection{Behaviour of the SF at the energy thresholds}

Following~\cite{CooperSimon_PRL_2015} and based on the microscopic insight discussed in the previous Sec.\ref{subsec:DSF_fine_structure}, we can identify the states corresponding to the energetic thresholds of $A_l(\omega)$: 
the lower threshold corresponds to the state labelled by the partition $[1^l]$, while the upper threshold corresponds to the state of partition $[2^{l/2}]$ if $l$ is even or $[2^{(l-1)/2},1]$ if $l$ is odd. These partitions have a simple interpretation in terms of the fermionic model: as in the DSF case, the state at the lower threshold of the SF has a deep-hole at $\sim -l$ (an impurity) off which particles at the Fermi point at $l=0$ (i.e. the Luttinger liquid) can scatter with small angular momentum exchanges.
On the other hand, the state at the upper threshold displays a pair of deep holes at $\sim -l/2$ and dramatically differs from the (higher energy) state with a high-energy particle at $\sim l$ related to the upper threshold of the DSF. Therefore, the upper threshold of the SF does not correspond to the highest energy eigenstate at the given angular momentum $l$.

\begin{figure}[t]
   	\begin{adjustbox}{width=.51\textwidth, totalheight=\textheight-2\baselineskip,keepaspectratio,right}
      	\includegraphics[]{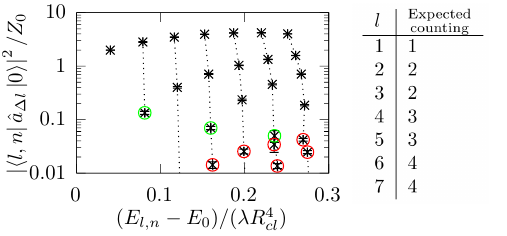}
    \end{adjustbox}
    \vspace{0.0cm}\caption{{Log-scale plot of the SF weight $|\bra{f}\hat a_{(N-1)/\nu-l}\ket{0}|^2$ (black points), normalized to the $l=0$ Laughlin-Laughlin matrix element $Z_0$, for $N=25$ bosons at $\nu=1/2$ in a quartic $\delta=4$ trap. 
    The data are the result of the full 2D numerical calculation.    
    The amplitudes relative to each  $l$-sector ($l=1\hdots7$) have been joined with black dashed lines as a guide for the eye.
    The point corresponding to $l=0$ is not shown as it equals $1$ by definition due to the normalization factor.
    The table shows the expected counting according to~\cite{CooperSimon_PRL_2015}; red circles correspond to points that should be excluded from the counting in order to agree with this counting prescription. Green circles indicate the least active point whose classification appears ambiguous.}
    \label{fig:sf_counting}}
\end{figure}

For small systems, our results are in agreement with exact diagonalization results. Thanks to the larger values of $N$ accessible to our calculations,
{ it is interesting to make a detailed comparison with the counting prescription derived from Haldane's fractional exclusion statistics principle~\cite{Haldane_PRL_1991,BernevigHaldane_PRL_2008} as discussed in~\cite{CooperSimon_PRL_2015}. 
As it is shown in Fig.~\ref{fig:sf_counting}, for odd values of $l$ the number of states with a significant SF weight indeed follows this counting prescription, while all other states (in red in the figure) have a much smaller SF weight.  For even $l$, instead, this classification is more ambiguous and may appear more accurate if we considered that one more state (in green in the figure) has a negligible SF weight.
A precursor of this fact can already be appreciated for the small system sizes considered in~\cite{CooperSimon_PRL_2015}: also in this work, for moderate $l$ excitations, the least active state systematically  has a smaller spectral weight for even $l$ than for odd $l$.} 
This different behaviour can be attributed to the qualitatively different structure of the state with the smallest possible SF weight at a given $l$, that is the state closest to the upper threshold, which corresponds to the $[2^{l/2}]$ and $[2^{(l-1)/2},1]$ partitions for respectively even and odd values of $l$. 
This unexpected suppression of the spectral weight is even more remarkable if one thinks the state of partition $[2^{l/2}]$ as the result of removing one particle of well-defined angular momentum $(N-1)/\nu-l$ from the $N$ particle Laughlin state partition, which is equivalent to the creation of a double quasihole at the same angular momentum. All other states with non-zero spectral weight correspond instead to the insertion of two quasiholes with distinct values of the angular momentum.

In analogy to the DSF shown in Fig.~\ref{fig:Energy_DSF}, also the SF { shown in Fig.~\ref{fig:spectralFunction}} displays a marked singularity at the lower threshold within each $l$ sector. The characterization of the functional form of the threshold behaviour however appears to be far more challenging than in the DSF case, and no robust conclusion can yet be drawn from the available theoretical insight and numerical data.

From the theoretical side, a study of the SF threshold within the bosonized theory is made difficult by the exponential form Eq.~\eqref{eq:bosonized_annihil} of the single-particle annihilation operator, to be contrasted to the expression \eqref{eq:dynamic_structure_factor} of the DSF that directly involves the density operator $\hat{\rho}$.
It is reasonable to think that the refermionized model may still be relevant for extracting information on the SF, especially at the lower energy threshold which, as we discussed above, corresponds to a simple situation in which a Luttinger liquid scatters from a deep-hole. However, a naive attempt to express the bosonized form of the particle annihilation operator Eq.~\eqref{eq:bosonized_annihil} in terms of the fermions $e^{-i\hat{\phi}(\theta)/\nu}=e^{-i\hat{\phi'}(\theta)/\sqrt{\nu}}\sim\Psi^{1/\sqrt{\nu}}$ leads to an ill-defined operator, whose value to characterize the SF singularities using a mobile-impurity model~\cite{ImambekovGlazman_RMP_2012} is far from obvious and calls for more sophisticated treatment. 

Serious difficulties are also present from the numerical side. Testing the power-law behaviour by fitting the SF $A_l(\omega)$ computed with the \nlchill model
requires working at even larger $l$ values than for the DSF case as the number of points that are available to the fit  for given $l$ is roughly halved as compared to the DSF case. Moving to higher $l$ values forces to work in larger Hilbert spaces leading to an intractable numerical complexity of the calculation.

In spite of these theoretical difficulties, the SF remains a quantity of key experimental interest. On one hand, it can be directly probed in the single-particle spectroscopy experiments proposed for FQH clouds of atoms or photons in~\cite{CooperSimon_PRL_2015,Umucalilar_PRA_2017}. Specially in the photonic case the SF is directly observable from the emission spectrum of the fluid, so our predictions are of direct application to the experiments.
On the other hand, a complete understanding of the SF will be instrumental to attack the harder questions related to the non-perturbative dynamics of the edge: large density dips at the edge of the FQH cloud can in fact be be produced by selectively removing particles close to the edge of the cloud. It is therefore very interesting to investigate the emerging non-linear Korteweg-de Vries hydrodynamics described by Eq.~\eqref{eq:NLXLL_hamiltonian}, which, analogously to what occurs at the classical level, may lead to shockwaves and solitons that chirally propagate  along the edge, a possibility already pointed out in~\cite{Bettelheim_PRL_2006,Wiegmann_PRL_2012}.

\section{Conclusions}\label{sec:conclusions}
In this work we have shown how the nonlinear chiral Luttinger liquid description of the low-energy dynamics of the edge modes of a fractional quantum Hall cloud introduced in~\cite{Nardin_PRA_2023} can be conveniently reformulated in terms of a model of massive and interacting chiral fermions in one-dimension.
Exploiting the physical insight offered by applying Tomonaga-Luttinger liquid techniques to the refermionized theory, we investigate the dynamic structure factor and the spectral function of the fractional quantum Hall fluid, providing quantitative explanations to the numerical predictions of full two-dimensional simulations of the fractional quantum Hall cloud.
In particular, our refermionized theory offers a physical interpretation to the dynamic structure factor broadening, which sets the edge-mode-decay time-scale. 
Building on established results for quantum impurity models in Luttinger liquids~\cite{ImambekovGlazman_RMP_2012}, we then investigate momentum-dependent power-law behaviour shown by the DSF close to its spectral edge:
the exponents turn out to be universal, for they only depend on the bulk filling fraction $\nu$ but are otherwise independent on the details of the confinement potential.

We then show how the nonlinear chiral Luttinger liquid model provides accurate information also on the spectral function describing the energy-dependence of the particle removal process. The quantitative agreement with the full 2D numerical calculation showcase the accuracy of the exponential form of the bosonized destruction operator within the nonlinear chiral Luttinger liquid theory.
Since the equation of motion of the edge-density operator has a Korteweg-de Vries form in the classical limit, and such an equation is well-known to admit solitonic solutions, a future task will be to address the emergent dynamics of the large density depletions caused by the removal of a particle close to the edge and investigate whether shock-wave-like behaviours can lead to the formation of solitons~\cite{Bettelheim_PRL_2006,Wiegmann_PRL_2012}.

Since our conclusions are based on a very generic model with short-ranged interactions, we anticipate that they straightforwardly apply to fractional quantum Hall fluids in either atomic or photonic synthetic quantum matter.
As such, they are ready to be experimentally verified with state-of-the-art technology and hold a great promise as a novel probe of the bulk topological order and its anyonic excitations. 
Future efforts will be devoted to the generalization of our approach to
the study of long-range interacting systems, to understand the interplay between the confinement-induced physics studied here and the non-vanishing interaction energy in a two-dimensional electron gas in solid-state devices.
Further steps will address exotic non-abelian quantum Hall states, such as the Moore-Read~\cite{MooreRead_NPB_1991}, which do host a more complex manifold of edge modes.
On a longer run, we believe our results will contribute paving the way towards the study of fractional quantum Hall fluids as a novel platform for nonlinear quantum optics of edge excitations with unprecedented dynamical and statistical properties.

\begin{acknowledgements}
    We acknowledge financial support from the Provincia Autonoma di Trento, from the Q@TN initiative, and from PNRR MUR project PE0000023-NQSTI.  Continuous illuminating discussions with Leonardo Mazza and Daniele De Bernardis are warmly acknowledged.
\end{acknowledgements}

\appendix

\section{Full 2D simulations: the numerical method}\label{appendix:MonteCarlo}
In this Appendix \ref{appendix:MonteCarlo} we briefly describe the numerical method used in the full two-dimensional calculations. While the general idea of the method was first introduced in~\cite{Nardin_PRA_2023}, here we made a few additions to it.

In the absence of any confinement potential, the Laughlin state Eq.~\eqref{eq_Laughlin} and its edge-excitations Eq.~\eqref{eq:laughlinEdge} are degenerate zero-energy eigenstates of specific model Hamiltonians with short-range interactions~\cite{Haldane_PRL_1983, TrugmanKivelson_PRB_1985, SimonRezayiCooper_PRB_2007_1,SimonRezayiCooper_PRB_2007_2}. When we include a cylindrically-symmetric confinement $V_\text{conf}(r)$ potential, the eigenstates of the system can be still labeled by an angular momentum quantum number $L=L_0+l$, where $L_0$ is the angular momentum of the Laughlin ground state and $l\geq0$ the additional angular momentum carried by the chiral edge excitation.
We furthermore consider the confinement to be weak compared to the interaction energy, $V_\text{conf}(R_{cl})\ll V_\text{int}$, and radially smooth, so that mixing with states above the many-body energy gap can be neglected. 

Under this simplifying assumption each many-body eigenstate $\ket{l,n}$ can be expanded over the Laughlin states basis as
\begin{equation}
    \ket{l,n}=\sum \sum_{\kappa_l} C_{\kappa_l}^{[l,n]} \frac{\ket{\kappa_l,l}}{\sqrt{\braket{\kappa_l,l|\kappa_l,l}}}
    \label{eq:app_eigenstate_expansion}
\end{equation}
where the states $\ket{\kappa_l,l}$ are a basis of linearly independent but not necessarily orthogonal states spanning the space of edge excitations { of the two-dimensional FQH cloud} at angular momentum $L_0+l$. 
These states are precisely those of Eq.~\eqref{eq:laughlinEdge}, 
\begin{equation}
\braket{\mathbf{r}_1,\hdots,\mathbf{r}_N|\kappa_l,l}= P_{\kappa_l}\left(\left\{z_i\right\}\right)\,\Psi_L(\left\{z_i\right\}).
\end{equation}
{{In particular, the index $\kappa_l$ runs through the integer partitions of $l$, namely 
through all the lists of positive integers $\kappa_l=[k_1,k_2,\dots]$, in decreasing order, such that $\sum_i k_i=l$, with the possibility of repetitions.}
To ensure the linear independence of the polynomials the partitions $\kappa_l$ have to be restricted to those containing at most $N$ elements: the total number of such states is $p_N(l)$; when $N\rightarrow\infty$, this number grows as $p_N(l\gg1)\sim\exp\left(\pi \sqrt{2l/3}\right) / \left(4l\sqrt{3}\right).$
}
In practice, the polynomial $P_{\kappa_l}\left(\left\{z_i\right\}\right)$ of Eq.~\eqref{eq:laughlinEdge} at a angular momentum  $l$ is written 
as a product of power-sum symmetric polynomials~\cite{SimonWavefunctionology_2020}{:
\begin{equation}
    P_{\kappa_l}(\{z_i\}) = \prod_{k \in \kappa_l} \left(\sum_{i=1}^N z_i^{k}\right).
\end{equation}
}
We remark that with such a choice the states $\ket{\kappa_l,l}$ are orthogonal only in the thermodynamic limit.
    
Projecting the many-body time-independent Schr\"odinger equation $H\ket{l,n}=E_{l,n}\ket{l,n}$ over these basis states, we obtain a Schr\"odinger equation in the form of a generalized eigenvalue problem for the expansion coefficients $C_{\kappa_l}^{[l,n]}$ appearing in Eq.~\eqref{eq:app_eigenstate_expansion},
\begin{equation}
    \label{eq:app_projectedSchrodinger}
        \mathds{H}_{\kappa_l,\kappa'_l}C_{\kappa'_l}^{[l,n]}
        =
        E_{l,n}\,\mathds{M}_{\kappa_l,\kappa'l} C_{\kappa'_l}^{[l,n]}\,.
\end{equation}
Here, the matrix elements $\mathds{H}$ and $\mathds{M}$ are defined as
\begin{equation}
    \label{eq:app_matrixElements}
    \begin{cases}
        \mathds{H}_{\kappa_l,\kappa'_l} = \frac{\bra{\kappa_l,l}H\ket{\kappa'_l,l}}{\sqrt{\braket{\kappa_l|\kappa_l}\braket{\kappa'_l|\kappa'_l}}}
        \\
        \mathds{M}_{\kappa_l,\kappa'_l} = \frac{\braket{\kappa_l,l|\kappa'_l,l}}{\sqrt{\braket{\kappa_l|\kappa_l}\braket{\kappa'_l|\kappa'_l}}}.
    \end{cases}
\end{equation}
Since the kinetic energy is constant within the lowest Landau level and the two-body interaction energy is assumed to be negligible within the subspace of Laughlin-like states (it is exactly zero for the case of contact-interacting bosons at $\nu=1/2$, or for specific short-ranged interactions at more general $\nu$), the effective Hamiltonian $\mathds{H}$ only includes the confinement potential $V_{\rm conf}(r)$, while the non-orthonormality of the basis wavefunctions is accounted for by the ``metric'' $\mathds{M}$.

The high-dimensional integrals in Eq.~\eqref{eq:app_matrixElements} are computed by means of a standard Metropolis-Monte Carlo sampling ~\cite{MetropolisUlam_JStat_1949,MetropolisTeller_JChemPhys_1953,Hastings_Biometrika_1970}. Since the states $\ket{\kappa_l,l}$ and $\ket{\kappa'_l,l}$ are polynomials of not-too-large degree with a Laughlin wavefunction Eq.~\eqref{eq_Laughlin} as common factor, they share most of their zeros, so the sampling procedure is highly efficient.

All other matrix elements, e.g. those appearing in the DSF and in the SF  are computed in an analogous way. This calculation is performed by first calculating the relevant matrix elements in the  basis $\ket{\kappa_l,l}$ of power-sum polynomials, and then rotating the results onto the basis of system eigenstates via the eigenvector matrix Eq.~\eqref{eq:app_eigenstate_expansion}. This eigenvector matrix thus requires to be computed with high accuracy, which typically requires a much larger number of Monte Carlo realizations than just obtaining the eigenvalues.

{
	\section{Bosonization of the Fermionized Hamiltonian}\label{appendix:bosonization}
	In this Appendix~\ref{appendix:bosonization} we briefly show that the fermionized model of Eq.~\eqref{eq:fermion_hamiltonian} and the bosonic model Eq.~\eqref{eq:NLXLL_hamiltonian_rescaled} are indeed equivalent.

    We begin by rewriting the free-part of the fermionic Hamiltonian Eq.~\eqref{eq:fermion_hamiltonian} in real space, using $\hat R_l^{\vphantom{\dagger}} = \frac{1}{\sqrt{2\pi}}\int e^{-i l \theta}\hat R^{\vphantom{\dagger}}(\theta)d\theta$.
    The linear part of the free-fermion term becomes
    \begin{equation}
        \label{eq:fermionic_linear}
        \begin{split}
        \hat H'_0 &= \left(\Omega -\frac{1}{2m^*}\right)  \sum_{l} l \hat R^\dagger_l \hat R^{\vphantom{\dagger}}_l =\\
        &=\left(\Omega -\frac{1}{2m^*}\right) \int \hat R^{\dagger}(\theta) \left(-i\partial_\theta\right) \hat R^{\vphantom{\dagger}}(\theta)d\theta ;
        \end{split}
    \end{equation}
    notice that we here included a $\propto 1/2m^*$ contribution to single out the purely quadratic part of the fermionic dispersion
    \begin{equation}
        \label{eq:fermionic_quadratic}
        \begin{split}
        \hat H'_1 &= \frac{1}{2m^*}\sum_{l}  l^2 \hat R^\dagger_l \hat R^{\vphantom{\dagger}}_l =\\
        &=\frac{1}{2m^*} \int \hat R^{\dagger}(\theta) \left(-\partial_\theta^2\right) \hat R^{\vphantom{\dagger}}(\theta)d\theta.
        \end{split}
    \end{equation}  

    When $\alpha=0$ and $m^*\rightarrow\infty$, the model is a single chiral branch of the Tomonaga-Luttinger model; we therefore introduce a Fermi-sea state Eq.~\eqref{eq:FermiSea} with all the states below $l\leq l_F=0$ occupied, and all the other states being empty. Notice that such a state has an infinite number of particles, but this is not problematic since we are only concerned with low-energy excitations around the Fermi surface at $l_F=0$. 
    We here always consider finite but ``small" values for both the interaction strength $\alpha$ and the reciprocal of the effective mass $(m^*)^{-1}$, 
    such that the state Eq.~\eqref{eq:FermiSea} is still the ground-state with respect to the long-wavelength $l/R_{\rm cl}$ excitations we are considering. For larger values of $l$, extra-terms in the Hamiltonian Eq.~\eqref{eq:NLXLL_hamiltonian} can be expected to appear~\cite{KunYang_PRL_2003, Dubail_PRB_2012,Fern_PRB_2018} - but they are irrelevant at the energy-scales we consider here.
    
    In this scenario, we can then bosonize this Hamiltonian by introducing a bosonic representation of the fermionic field operator~\cite{Giamarchi_QP1D_2004} $\hat R(\theta) = \sqrt{\frac{1}{2\pi\epsilon}} e^{-i \hat \phi'}$,
    where the mode decomposition of the operator $\hat\phi'$ in terms of bosonic creation/annihilation operators $[\hat a_{l'}^{\vphantom{\dagger}},\hat a_l^{\dagger}]=\delta_{l,l'}$ reads $\hat \phi' = \sum_{l>0} \frac{i}{\sqrt{l}}\left(e^{i l \theta} \hat a_l^{\vphantom{\dagger}} - e^{-i l \theta} \hat a_l^{\dagger}\right) e^{-\frac{\epsilon}{2}\,l}$, and $\epsilon\rightarrow0$ being a length-scale which serves as an ultraviolet cutoff.
    The density operator $\hat\rho'$ is written as $\hat\rho' = -\frac{1}{2\pi}\partial_\theta\hat\phi'$.
    Notice that this operator does satisfy the commutation relations Eq.~\eqref{eq:KacMoodyCommutator_rescaled}.
    It can also be checked that $\lim_{\theta'\rightarrow\theta}\left(:\!\hat R^\dagger(\theta')\hat R(\theta)\!:\right)_{|\theta'-\theta|\gg\epsilon} = \hat\rho(\theta)$, where the colons denote normal ordering with respect to the vacuum of bosonic excitations.
    Namely, in order to obtain a well defined expression for the density operator, the field operator product is performed at different positions, $\hat R^\dagger(\theta')\hat R(\theta)$, the result normally ordered and expanded for distances much larger than the cutoff length-scale $\epsilon$, and finally the limit $\theta'\rightarrow\theta$ is taken.
    In the same way, the linear part of the fermionic dispersion Eq.~\eqref{eq:fermionic_linear} gets regularized as  $\lim_{\theta'\rightarrow\theta}\left(:\!\hat R^\dagger(\theta')(-i\partial_\theta)\hat R(\theta)\!:\right)_{|\theta'-\theta|\gg\epsilon}$, yielding the standard result~\cite{Giamarchi_QP1D_2004}
    \begin{equation}
        \hat H_0' = \pi \left(\Omega -\frac{1}{2m^*}\right) \int :\!{\hat\rho}'^2(\theta)\!:d\theta.
    \end{equation}
    The quadratic term Eq.~\eqref{eq:fermionic_quadratic} can be regularized in the same way. After a lengthy calculation we find
    \begin{equation}
        \hat H_1' = \frac{1}{2m^*} \int \left(\frac{4\pi^2}{3}:\!{\hat\rho}'^3(\theta)\!:+\,\pi:\!{\hat\rho}'^2(\theta)\!:\right)d\theta.
    \end{equation}
    Notice how a quadratic term $\propto \,:\!\!{\hat\rho}'^2(\theta)\!\!:$ appears, which however cancels out the identical term in $\hat H_0$ when adding the two results up
    \begin{equation}
        \hat H_0'+\hat H_1' = \int\left( \pi \Omega :\!{\hat\rho}'^2(\theta)\!: +  \frac{2\pi^2}{3m^*}:\!{\hat\rho}'^3\!:\right)d\theta.
    \end{equation}    
    
    Finally, the fermionic interaction term is straightforward since it comes already in the form of a density-density interaction~\cite{Giamarchi_QP1D_2004}; using $\hat\rho_l'=\int e^{-i l \theta}\hat\rho'(\theta)d\theta$ we get
    \begin{equation}
        \hat H'_2 =  - \frac{\alpha}{2}\sum_l l^2 \hat{\rho}_{-l}{\!\!\!\!\!'\,\,\,\,}\hat{\rho}_{l}{\!'} = -\pi\alpha\int d\theta :\!\left(\partial_\theta\hat{\rho}'\right)^2\!:.  
    \end{equation}  
    putting all the results together yield indeed Eq.~\eqref{eq:NLXLL_hamiltonian_rescaled}.
}

\section{DSF broadening in a quartic  trap}\label{appendix:broadening}
\begin{figure}[htbp]
   	\begin{adjustbox}{width=.5\textwidth, totalheight=\textheight-2\baselineskip,keepaspectratio,right}
      	\includegraphics[]{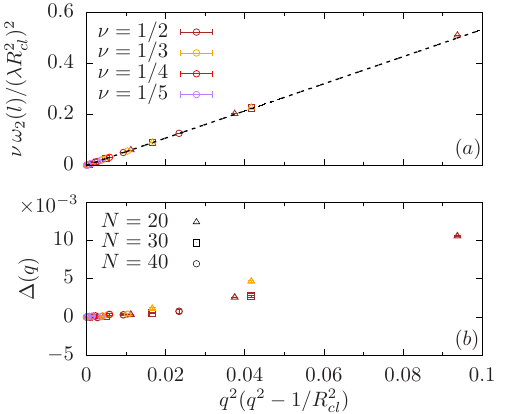}
    \end{adjustbox}
    \vspace{0.0cm}\caption{(a) Plot of the (normalized) second moment $\omega_2$ of the DSF in a quartic $\delta=4$ trap, for different system sizes $N$ and filling fractions $\nu$. In order to highlight the long-wavelength limiting behaviour, we plot $\omega_2$ as a function of $q^2(q^2-1/R_{cl}^2)$ rather than $l^2(l^2-1)$ where $q$ is the physical wavevector $q=l/R_{cl}$ of the edge excitation. The result of the full 2D numerical calculation (symbols) is compared with the analytical prediction Eq.~\eqref{eq:broadening2} of the non-interacting fermion approximation (black-dashed line). (b) Difference between the numerical and the analytical results shown in (a): the difference tends to zero in the $q\to 0$ long wavelength limit. 
    \label{fig:quarticTrapBroadening}}
\end{figure}

In this Appendix we present a more detailed analysis of the second moment
\begin{equation}
    \omega_2(l) = \frac{\int (\omega - \omega_1(l))^2 S_l(\omega) d\omega}{\int S_l(\omega) d\omega}
\end{equation}
 of the edge DSF in the case of a weak quartic $V_\text{conf}=\lambda r^4$ potential confining the FQH cloud and we compare the result to the theoretical prediction of Eq.~\eqref{eq:broadening2}.
This case is particularly simple: in the quartic trap, the angular velocity gradient (the second derivative with respect to the angular momentum $\partial_l = \frac{\partial r}{\partial l}\,\partial_r\approx r^{-1}\partial_r$) at the position of the cloud's edge is in fact constant and position independent [Eq.~\eqref{eq:curvature}], 
\begin{equation}
c=\left.(r^{-1}\partial_r)^2\,V_\text{conf}\right|_{r=R_{cl}}=8\lambda    
\end{equation}
so we have [Eq.~\eqref{eq:broadening2}]
\begin{equation}
    \label{eq:appendixA_broadening}
    \omega_2(l) = \frac{16\lambda^2}{3\nu}\,l^2(l^2-1)\,.
\end{equation}

When the product $\nu\times \omega_2(l)$ is plotted, data from different filling factors $\nu$ are expected to collapse on the same curve.
In order to better highlight the long-wavelength limit, data for different values of the particle number $N$ and filling factor $\nu$ are plotted in Fig.~\ref{fig:quarticTrapBroadening} in terms of $q^2(q^2-R_{cl}^{-2})$, where $q=l/R_{cl}$ is the physical wavevector of the edge excitation: it is apparent how all points accurately fall on the expected straight line Eq.~\eqref{eq:appendixA_broadening}. The difference $\Delta(q)$ from this line are shown in the bottom panel: the difference is always small and tends to zero in the long-wavelength limit.

\begin{figure}[htbp]
	\begin{adjustbox}{width=.5\textwidth, totalheight=\textheight-2\baselineskip,keepaspectratio,right}
		\includegraphics[]{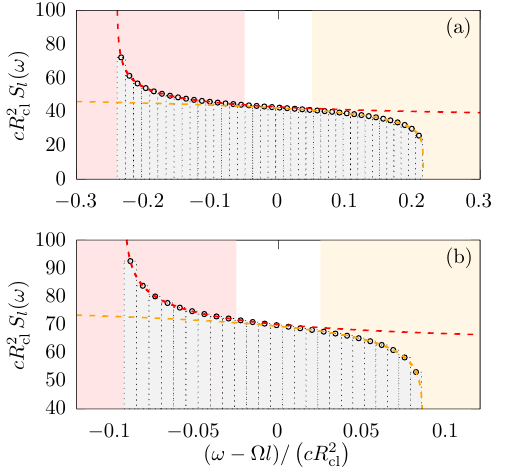}
	\end{adjustbox}
	\vspace{0.0cm}\caption{
		{ Plot of the DSF $S_l(\omega)$ at (a) $l=40$ and (b) $l=25$, obtained through numerical exact diagonalization of the \nlchill Hamiltonian Eq.~\eqref{eq:NLXLL_hamiltonian}, versus the energy $\omega-\Omega l$ rescaled to the effective curvature parameter $c R_{\rm cl}^2$. 
		Only the $l$ largest DSF weights (black points) are shown here; the small contributions coming from the other states (see e.g. Fig.~\ref{fig:DSFLogScale}) are not shown since not important to the current discussion. The grey boxes correspond to smearing of the $\delta$-function peaks (see discussion in Appendix~\ref{appendix:dsf_fitting}).
		The red/orange curves are fits of Eq.~\eqref{eq:dsf_fit} against the DSF in the region highlighted by red/orange shading.}}
		\label{fig:dsf_fit}
\end{figure}

{
\section{Fitting the power-law behavior at the DSF thresholds}\label{appendix:dsf_fitting}
In this Appendix we briefly comment on how we numerically extracted the DSF threshold exponents displayed in Fig.~\ref{fig:DSFexponents} from the \nlchill DSF, obtained by numerical exact diagonalization of Eq.~\eqref{eq:NLXLL_hamiltonian}.

Instead of directly fitting Eq.~\eqref{eq:dsf_threshold_behavior} to the data, we allowed for the presence of logarithmic corrections~\cite{Pustilnik_PRL_2006}
\begin{equation}
	\label{eq:dsf_fit}
	\begin{cases}
		S_l(\omega\sim\omega_-) = S_-\,\theta(\omega-\omega_-)\left(\frac{1}{\omega-\omega_-} + \delta_-\right)^{\mu_-}\\
		S_l(\omega\sim\omega_+) = S_+\,\theta(\omega_+-\omega)\left(\frac{1}{\omega_+-\omega}+\delta_+\right)^{-\mu_+};
	\end{cases}
\end{equation}
when $\omega\simeq\omega_{\pm}$, the logarithmic corrections $\delta_{\pm}$ are irrelevant, but it is important to keep them into account when dealing with finite portions of the DSF at finite transferred angular momentum $l$.
We fitted $S_l(\omega_{\pm})$ close to the threshold $\omega_{\pm}$, with $S_\pm$, $\omega_{\pm}$, $\delta_{\pm}$ and $\mu_{\pm}$ kept as free parameters.

An example of a fit at $N=5000$, $l=40$ is shown in Fig.~\ref{fig:dsf_fit}(a).
We here remark that we did not perform the fits directly against the DSF weights obtained through the exact-diagonalization of the \nlchill model of Eq.~\eqref{eq:NLXLL_hamiltonian}; rather, we first broadened the $\delta$-peaks (see Eq.~\eqref{eq:dynamic_structure_factor}) into boxes so as to reproduce $S_l(\omega)$, as shown in Fig.~\ref{fig:dsf_fit}. 
Notice how this slightly modifies the relative weights if the energy-spacing between adjacent points is non-uniform.
We then fitted Eq.~\eqref{eq:dsf_fit} against the central point (black circles) within the regions highlighted by semi-transparent rectangles.
It can be seen how this simple ansatz correctly captures the DSF close to the thresholds $\omega_\pm$.

As a final remark, it is clear that the fitting procedure becomes problematic for small values of $l$, since there are not enough degrees of freedom to fit Eq.~\eqref{eq:dsf_fit} meaningfully; the smallest angular-momenta at which we have been able to fit the DSF meaningfully was $l=25$; we show an example in Fig.~\ref{fig:dsf_fit}(b).
}

\section{Overlaps with Jack polynomials}\label{appendix:Jacks}
\begin{figure}[htbp]
   	\begin{adjustbox}{width=.5\textwidth, totalheight=\textheight-2\baselineskip,keepaspectratio,right}
      	\includegraphics[]{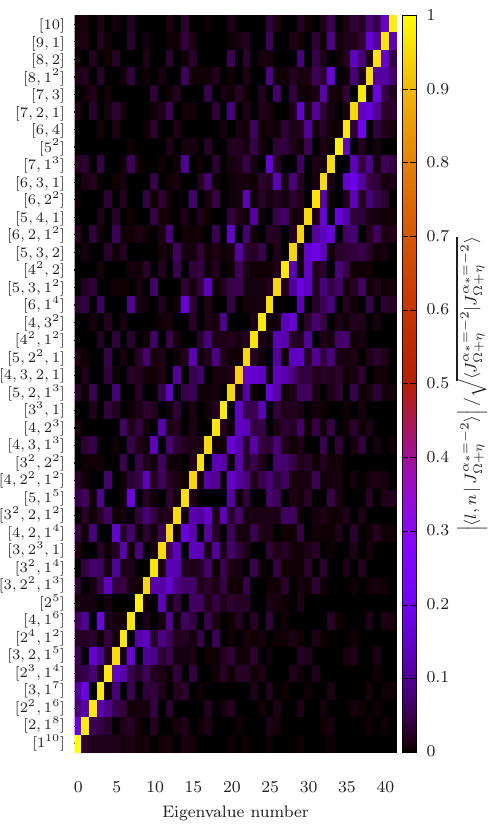}
    \end{adjustbox}
    \vspace{0.0cm}\caption{
    Overlap matrix between the numerically obtained eigenstates $\ket{l,n}$ for a FQH cloud at $\nu=1/2$ in a quartic trap $V_\text{conf}=\lambda r^4$ and the (normalized) Jack polynomials $J_{\Omega+\eta}^{\alpha_*}$ with $\alpha_*=-2/(r-1)$ and $r=1/\nu$. 
     On the $x$ axis, the system's eigenvectors $\ket{l,n}$ are ordered by increasing eigenvalue number $n$. 
     On the $y$ axis we report the Jack edge-partitions $\eta$ with which the overlap has been computed. The edge-partitions $\eta$, on the $y$-axis, have been ordered in such a way that the the largest overlap (close to $1$) lies on the diagonal.
    \label{fig:quarticTrapJacksOverlaps}}
\end{figure}

In this Appendix we explicitly show how the eigenmodes at filling $\nu=1/2$ in the quartic trap have large overlaps with a single Jack polynomial~\cite{LeeHuWan_PRB_2014,Macaluso_PRA_2017,Macaluso_PRA_2018} of parameter $\alpha_*=-2/(r-1)$ with $r=1/\nu=2$ and labelled by a $(1,r)$-admissible root-configuration $\Omega'$. 

We expand $\Omega' = \Omega+\eta$, where $\Omega$ is the Laughlin state root partition~\cite{BernevigHaldane_PRL_2008} and $\eta$ is the \textit{edge-partition}, labelling a basis of edge excitations at angular momentum $l$. As described in~\cite{LeeHuWan_PRB_2014}, analogously to Eq.~\eqref{eq:laughlinEdge}, we then have $J_{\Omega+\eta}^{\alpha_*}=J_{\eta}^{\beta_*}J_{\Omega}^{\alpha_*}$ where $\beta_*=2/(r+1)$ and $J_{\Omega}^{\alpha_*}$ is the Laughlin state at filling $1/r=1/2$ in our case.

Since the Laughlin state factors out, the overlaps between the eigenstates $\ket{l,n}$ and the  edge-Jacks $J_{\Omega'}^{\alpha_*}$ are easily computed by means of a generalization of the Monte Carlo sampling discussed above. We start by expanding the eigenstates at angular momentum $l$ as in Eq.~\eqref{eq:app_eigenstate_expansion} $\ket{l,n}=\sum_{\kappa_l} C_{\kappa_l}^{[l,n]} \frac{\ket{\kappa_l,l}}{\sqrt{\braket{\kappa_l,l|\kappa_l,l}}}$, so that we can express the properly-normalized matrix elements of interest as
\begin{equation}
    \label{eq:app_overlaps_1}
    \frac{\braket{l,n|J_{\Omega'}^{-2}}}{\sqrt{\braket{J_{\Omega'}^{-2}|J_{\Omega'}^{-2}}}}
    =
    \sum_{\kappa_l} C_{\kappa_l}^{[l,n]*}\,
    \underbrace{\frac{\braket{\kappa_l,l|J_{\Omega'}^{-2}}   }{\sqrt{\braket{\kappa_l,l|\kappa_l,l}\braket{J_{\Omega'}^{-2}|J_{\Omega'}^{-2}}}}}_{O_{\kappa_l;\eta}}.
\end{equation}
Introducing the normalized probability density function
\begin{equation}
    \mathcal P_{\kappa_l}(z) = \frac{|P_{\kappa_l}(z) \Psi_L(z)|^2}{\int |P_{\kappa_l}(z) \Psi_L(z)|^2\mathcal D z}
\end{equation}
where $z$ is a shorthand for $z_1,\hdots,z_N$, $\mathcal D z=d^2z_1\hdots d^2z_N$ and $\braket{z|\kappa_l,l}=P_{\kappa_l}(z) \Psi_L(z)$, the normalized overlaps $O_{\kappa_l;\eta}$ appearing on the right-hand side of Eq.~\eqref{eq:app_overlaps_1} read
\begin{equation}
    O_{\kappa_l;\eta} = \frac{\int \mathcal P_{\kappa_l}(z)\frac{J_\eta^{\beta_*}(z)}{P_{\kappa_l}(z)}\mathcal Dz}{\sqrt{\int \mathcal P_{\kappa_l}(z) \left|\frac{J_\eta^{\beta_*}(z)}{P_{\kappa_l}(z)}\right|^2\mathcal Dz}}\,.
\end{equation}
These can be computed through Monte Carlo sampling: once again, since the states $J_{\Omega+\eta}^{\alpha_*}$ and $P_{\gamma_l}\Psi_L$ are polynomials of not-too-large degree with a Laughlin wavefunction Eq.~\eqref{eq_Laughlin} as common factor, they share most of their zeros and the sampling procedure is highly efficient.
Finally, the Jack polynomials $J_\eta^{\beta_*}$ are computed by expanding them in the basis of power-sum symmetric polynomials seen above, $J_\eta^{\beta_*}(z)=\sum_{\gamma_l} j_{\eta}^{[\gamma_l]}\,P_{\gamma_l}(z)$.

\begin{figure}[ht!]
	\begin{adjustbox}{width=.5\textwidth, totalheight=\textheight-2\baselineskip,keepaspectratio,right}
		\includegraphics[]{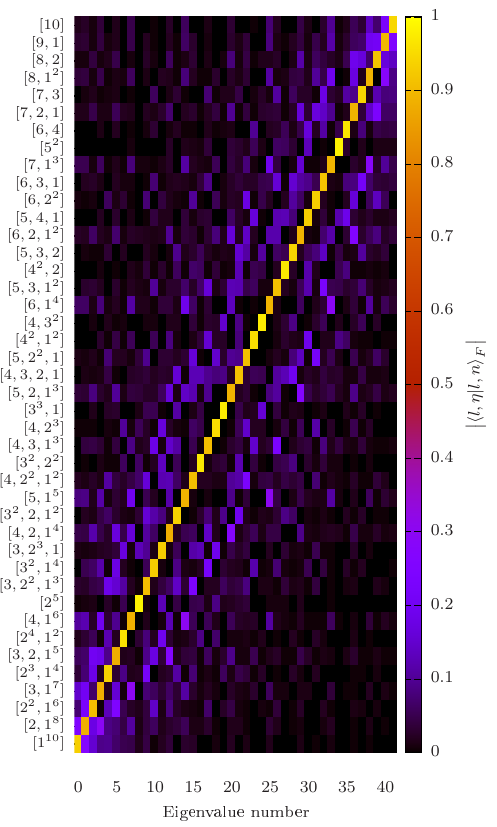}
	\end{adjustbox}
	\vspace{0.0cm}\caption{Overlap matrix between the numerically obtained eigenstates $\ket{l,n}_F$ of the fermionic model [Eq.~\eqref{eq:fermion_hamiltonian}] and the non-interacting particle-hole states labelled by the partition $\eta$ shown on the $y$ axis.
		On the $x$ axis, the states are ordered by increasing $n$. The partitions $\eta$ on the $y$ axis have been ordered in the same way as in Fig.\ref{fig:quarticTrapJacksOverlaps}. Note how this choice leads to the largest overlap being on the main diagonal.
		\label{fig:FermionOverlaps}}
\end{figure}

The results for the overlaps matrix are shown in Fig.~\ref{fig:quarticTrapJacksOverlaps}. On the $x$ axis, the system's eigenvectors $\ket{l,n}$ are ordered by increasing eigenvalue number $n$ (which is, by increasing energy eigenvalue). On the $y$ axis we show the edge-partition $\eta$ labelling a specific Jack; we ordered the partitions along the $y$-axis so that the largest overlap lies on the diagonal.
It is apparent how the eigenstates of our system have a large overlap ($\gtrsim0.95$) with a single Jack polynomial, while the largest off-diagonal component remains $\lesssim0.2$.

Finally, we want to compare the partitions $\eta$ labelling the Jacks with those labelling particle-hole excitations in the refermionized model [Eq.~\eqref{eq:fermion_hamiltonian}]. In order to do so,
in Fig.~\ref{fig:FermionOverlaps} we show 
the overlaps between the eigenstates $\ket{l,n}_F$ of the fermionic model with the non-interacting particle-hole excitations labelled by a partition $\eta$ and carrying the same angular momentum $l$, $\ket{l,\eta}$.
On the $x$ axis, the system's eigenvectors $\ket{l,n}_F$ are ordered by increasing eigenvalue number $n$ (which is, by increasing energy eigenvalue).
On the $y$ axis we show the partitions $\eta$ labelling the particle-hole states; the same ordering is adopted as in Fig.~\ref{fig:quarticTrapJacksOverlaps}.
It is apparent from Fig.~\ref{fig:FermionOverlaps} how each eigenstate of the refermionized model has a dominant ($\gtrsim90\%$) overlap with a single non-interacting particle-hole state.
Remarkably, the fact that the weight is concentrated on the main diagonal of the overlap matrix, together with the energies of the \nlchill being in good agreement with those of the full 2D FQH,
shows that the partition $\eta$ labelling the fermionic state through a non-interacting particle-hole state which satisfies the Pauli exclusion principle
is the same as the one labelling the full 2D FQH eigenstates through a Jack polynomial satisfying a generalized exclusion principle~\cite{BernevigHaldane_PRL_2008}.



\section{Monte Carlo calculation of the matrix elements of the spectral function}\label{appendix:SpectralFunctionMC}
In this Appendix we briefly describe how the matrix elements appearing in the SF Eq.~\eqref{eq:spectralFunction} can be written in a way which is amenable of Monte Carlo sampling.
We expand the annihilation operators at angular momentum $\Delta l = (N-1)/\nu-l$  as 
\begin{equation}
\hat{a}_{\Delta l} = \int dz_1 \Phi_{\Delta l}^*(z_1) \hat{\psi}(z_1),
\end{equation}
where
$\Phi_{\Delta l}(z_1)$ are the single-particle lowest Landau level orbitals in the circular gauge $\Phi_{\Delta l}(z_1) = \left(z_1/\sqrt{2}\right)^{\Delta l} \exp(-|z_1|^2/4) /\sqrt{2\pi \Delta l!}$.
We then write the matrix elements explicitly in first quantization language as
\begin{equation}
    \label{eq:spectralFunctionMicroscopicMC}
    \begin{split}
    \bra{f}\hat{a}_{\Delta l}\ket{0} 
    =&\sqrt{N}\int \left(\frac{{\Psi_{f}^{[N-1]}}(z_2\ldots z_N)}{\mathcal{N}_f}\right)^* \times \\\times&\frac{\Psi_0^{[N]}(z_1\ldots z_N)}{\mathcal{N}_0}\, \Phi_{\Delta l}^*(z_1) \prod_{i=1}^N  dz_i
    \end{split}
\end{equation}
where we defined the normalization factors as
\begin{equation}
    \mathcal{N}_0=\left[\int \left|\Psi_0^{[N]}(z_1\ldots z_N)\right|^2 \prod\nolimits_{i=1}^N  dz_i\right]^{1/2}
\end{equation}
and
\begin{equation}
    \mathcal{N}_f=\left[\int \left|\Psi_f^{[N-1]}(z_2\ldots z_N) \prod\nolimits_{i=2}^N  dz_i\right|^2\right]^{1/2}.
\end{equation}
The final states $\Psi_{f}$ are expanded as power-sum symmetric polynomials multiplying a $(N-1)$-particles Laughlin wavefunction  $\Psi_L^{[N-1]}(z_2,\ldots,z_N)$. 
We find that the Monte Carlo sampling of the integrals is strongly facilitated by introducing the probability distribution function
\begin{equation}
    \label{eq:PDF}
    \begin{split}
    \mathcal{P}&(z_1\ldots z_N) =\\&= \frac{|\Psi_0^{[N]}(z_1\hdots z_N)\Psi_0^{[N-1]}(z_2\hdots z_N)|}{\int |\Psi_0^{[N]}(z_1\hdots z_N)\Psi_0^{[N-1]}(z_2\hdots z_N)|\prod_{i=1}^N  dz_i}
    \end{split}
\end{equation}
since in this way we have zeros of the correct order at each particle's position, which makes the sampling more effective. 
The obvious disadvantage is that, since not all the coordinates are treated on an equal footing, the relevant matrix element which needs to be sampled [Eq.~\eqref{eq:spectralFunctionMicroscopicMC}]
cannot be symmetrized over all the coordinates: we however empirically find the usage of Eq.~\eqref{eq:PDF} to be a compromise worth the additional computational price.

{
	\section{Numerical calculation for the matrix elements of the spectral function within the \nlchill model}\label{appendix:sf}
	In this Appendix we further describe how the matrix elements of the annihilation operator Eq.~\eqref{eq:annihilation_operator} have been calculated within the \nlchill model of Eq.~\eqref{eq:NLXLL_hamiltonian}.
	We begin by rewriting Eq.~\eqref{eq:bosonized_annihil} in terms of the mode expansion of the density operator $\hat \rho(\theta) = \frac{1}{2\pi}\sum_{l>0}\sqrt{l\nu}\left(e^{i l \theta}\hat b_l^{\vphantom{\dagger}} + e^{-i l \theta} \hat b^{\dagger}_l\right)$, with $[\hat b_l^{\vphantom{\dagger}},\hat b^{\dagger}_{l'}]=\delta_{l,l'}$, as
	\begin{equation}
		\hat\psi(\theta)=\exp\left[\sum_{l>0}\left(\alpha_l(\theta) \hat b_l^{\dagger} - \alpha_l^*(\theta) \hat b_l^{\vphantom{\dagger}}\right)\right].
	\end{equation}
	Here $\alpha_l(\theta) = -\frac{e^{-i l \theta}}{\sqrt{\nu\,l}}$, and the Klein factor~\cite{Giamarchi_QP1D_2004} has been omitted, since it does not contribute to the matrix elements we are interested in~\cite{Jolad_PRB_2007,Jolad_PRL_2009,Jolad_PRB_2010}.
	It can be checked that $\bra{0}\hat\psi(\theta)^\dagger \hat\rho(\theta')\hat\psi(\theta)\ket{0} = -\delta(\theta-\theta')$; namely, in the state $\hat\psi(\theta)\ket{0}$ a unit charge has been displaced away from position $\theta$ from the ground-state $\ket{0}$~\cite{Giamarchi_QP1D_2004}.
	
	Within our exact-diagonalization code, the eigenstates $\ket{l,n}$ of the \nlchill Hamiltonian Eq.~\eqref{eq:NLXLL_hamiltonian} are expanded onto bosonic Fock states $\ket{l,F}=\prod_{l>0}'\frac{\left(\hat b^\dagger_l\right)^{n_l(F)}}{\sqrt{n_l(F)!}}\ket{0}$ which carry well defined angular momentum $\sum_{l'}l' n_{l'}(F)=l$. Here the primed product is a reminder for this constraint.
	In order to compute the matrix elements $\bra{l,n} \hat\psi(\theta)\ket{0}$ we therefore need the Fock-state decomposition of $\hat\psi(\theta)\ket{0}$; using the Baker–Campbell–Hausdorff formula this can rewritten as
	\begin{equation}
		\begin{aligned}
		\hat\psi(\theta)\ket{0}=&\left(\prod_{l>0}e^{-\frac{1}{2}|\alpha_l(\theta)|^2}\right) \prod_{l>0}e^{\alpha_l(\theta)\hat b_l^\dagger}\ket{0}=\\=&
		\left(\prod_{l>0}e^{-\frac{1}{2}|\alpha_l(\theta)|^2}\right) \prod_{l>0}\sum_{n=0}^\infty\frac{\alpha_l^n(\theta)}{n!} \left(\hat b_l^\dagger\right)^n\ket{0}.
		\end{aligned}
	\end{equation}
	It is not difficult to realize that the last term is a summation over all the possible Fock states and that it can be rewritten as
	\begin{equation}
		\hat\psi(\theta)\ket{0}=
		\left(\prod_{l>0}e^{-\frac{1}{2}|\alpha_l(\theta)|^2}\right) \sum_{F}\prod_{l>0}\frac{\alpha_l^{n_l(F)}(\theta)}{\sqrt{n_l(F)!}} \ket{F}
	\end{equation}
	and therefore
	\begin{equation}
        \label{eq:lF}
		\bra{l,F}\hat\psi(\theta)\ket{0}=
		\left(\prod_{l>0}e^{-\frac{1}{2}|\alpha_l(\theta)|^2}\right) \smash{\prod_{l>0}}\raisebox{4pt}{$'$}\frac{\alpha_l^{n_l(F)}(\theta)}{\sqrt{n_l(F)!}}.
	\end{equation}		
	Notice here that while the normalization part $\sqrt{Z_0}=\prod_{l>0}e^{-\frac{1}{2}|\alpha_l(\theta)|^2}$ is ultraviolet divergent and, thus, cutoff dependent, the remaining part of the expression is well-defined. Moreover, the ground-state expectation value $\bra{0'}\hat\psi(\theta)\ket{0}=\sqrt{Z_0}$ (here $\ket{0'}$ is still the vacuum of bosonic excitations, but with one particle less) gives exactly this normalization factor, which can thus be isolated within Eq.~\eqref{eq:lF} leading to an unambiguous result for the final factor of interest.
	
	Finally, given the constraint $\sum_{l'}l' n_{l'}(F)=l$ the $\theta$ dependence can be extracted
	\begin{equation}
	\frac{\bra{l,F}\hat\psi(\theta)\ket{0}}{\sqrt{Z_0}}=
	 e^{-i l \theta}\,\smash{\prod_{l>0}}\raisebox{4pt}{$'$}\frac{\left(-1\right)^{n_l(F)}}{\sqrt{(\nu l)^{n_l(F)}n_l(F)!}},
	\end{equation}
	which is the same expression quoted in Ref.~\cite{Jolad_PRB_2010}.
	The matrix elements $Z_0^{-1}\,|\bra{l,n}\hat a_{\Delta l}\ket{0}|^2$ in the basis of eigenstates $\ket{l,n}$ we were looking for,    
    can finally be computed by plugging the previous expression in Eq.~\eqref{eq:annihilation_operator} and rotating onto the basis of eigenvectors of the \nlchill model Eq.~\eqref{eq:NLXLL_hamiltonian}, which we determine by numerical diagonalization.
}

\bibliographystyle{apsrev4-2}
\bibliography{FQHDynamics.bib}

@article{ImambekovGlazman_RMP_2012,
  title = {One-dimensional quantum liquids: Beyond the Luttinger liquid paradigm},
  author = {Imambekov, Adilet and Schmidt, Thomas L. and Glazman, Leonid I.},
  journal = {Rev. Mod. Phys.},
  volume = {84},
  issue = {3},
  pages = {1253--1306},
  numpages = {0},
  year = {2012},
  month = {Sep},
  publisher = {American Physical Society},
  doi = {10.1103/RevModPhys.84.1253},
  url = {https://link.aps.org/doi/10.1103/RevModPhys.84.1253}
}

@article{Pustilnik_PRL_2006,
  title = {Dynamic Response of One-Dimensional Interacting Fermions},
  author = {Pustilnik, M. and Khodas, M. and Kamenev, A. and Glazman, L. I.},
  journal = {Phys. Rev. Lett.},
  volume = {96},
  issue = {19},
  pages = {196405},
  numpages = {4},
  year = {2006},
  month = {May},
  publisher = {American Physical Society},
  doi = {10.1103/PhysRevLett.96.196405},
  url = {https://link.aps.org/doi/10.1103/PhysRevLett.96.196405}
}

@article{Macaluso_PRA_2018,
  title = {Ring-shaped fractional quantum Hall liquids with hard-wall potentials},
  author = {Macaluso, E. and Carusotto, I.},
  journal = {Phys. Rev. A},
  volume = {98},
  issue = {1},
  pages = {013605},
  numpages = {17},
  year = {2018},
  month = {Jul},
  publisher = {American Physical Society},
  doi = {10.1103/PhysRevA.98.013605},
  url = {https://link.aps.org/doi/10.1103/PhysRevA.98.013605}
}

@article{Macaluso_PRA_2017,
  title = {Hard-wall confinement of a fractional quantum Hall liquid},
  author = {Macaluso, E. and Carusotto, I.},
  journal = {Phys. Rev. A},
  volume = {96},
  issue = {4},
  pages = {043607},
  numpages = {14},
  year = {2017},
  month = {Oct},
  publisher = {American Physical Society},
  doi = {10.1103/PhysRevA.96.043607},
  url = {https://link.aps.org/doi/10.1103/PhysRevA.96.043607}
}

@article{CooperSimon_PRL_2015,
  title = {Signatures of Fractional Exclusion Statistics in the Spectroscopy of Quantum Hall Droplets},
  author = {Cooper, Nigel R. and Simon, Steven H.},
  journal = {Phys. Rev. Lett.},
  volume = {114},
  issue = {10},
  pages = {106802},
  numpages = {5},
  year = {2015},
  month = {Mar},
  publisher = {American Physical Society},
  doi = {10.1103/PhysRevLett.114.106802},
  url = {https://link.aps.org/doi/10.1103/PhysRevLett.114.106802}
}

@book{Giamarchi_QP1D_2004,
  title={Quantum Physics in One Dimension},
  author={Giamarchi, T. and Oxford University Press},
  isbn={9780198525004},
  lccn={2004299020},
  series={International Series of Monographs on Physics},
  url={https://books.google.it/books?id=1MwTDAAAQBAJ},
  year={2004},
  publisher={Clarendon Press}
}

@article{Wen_intJModPhysB_1992,
  title={Theory of the Edge States in Fractional Quantum Hall Effects},
  author={Xiao-Gang Wen},
  journal={International Journal of Modern Physics B},
  year={1992},
  volume={06},
  pages={1711-1762}
}

@article{Haldane_JPC_1981,
doi = {10.1088/0022-3719/14/19/010},
url = {https://dx.doi.org/10.1088/0022-3719/14/19/010},
year = {1981},
month = {jul},
publisher = {},
volume = {14},
number = {19},
pages = {2585},
author = {F D M Haldane},
title = {'Luttinger liquid theory' of one-dimensional quantum fluids. I. Properties of the Luttinger model and their extension to the general 1D interacting spinless Fermi gas},
journal = {Journal of Physics C: Solid State Physics}
}

@article{Chang_PRL_1996,
  title = {Observation of Chiral Luttinger Behavior in Electron Tunneling into Fractional Quantum Hall Edges},
  author = {Chang, A. M. and Pfeiffer, L. N. and West, K. W.},
  journal = {Phys. Rev. Lett.},
  volume = {77},
  issue = {12},
  pages = {2538--2541},
  numpages = {0},
  year = {1996},
  month = {Sep},
  publisher = {American Physical Society},
  doi = {10.1103/PhysRevLett.77.2538},
  url = {https://link.aps.org/doi/10.1103/PhysRevLett.77.2538}
}

@article{PalaciosMacDonald_PRL_1996,
  title = {Numerical Tests of the Chiral Luttinger Liquid Theory for Fractional Hall Edges},
  author = {Palacios, J. J. and MacDonald, A. H.},
  journal = {Phys. Rev. Lett.},
  volume = {76},
  issue = {1},
  pages = {118--121},
  numpages = {0},
  year = {1996},
  month = {Jan},
  publisher = {American Physical Society},
  doi = {10.1103/PhysRevLett.76.118},
  url = {https://link.aps.org/doi/10.1103/PhysRevLett.76.118}
}

@article{Wan_PRB_2003,
  title = {Edge reconstruction in the fractional quantum Hall regime},
  author = {Wan, Xin and Rezayi, E. H. and Yang, Kun},
  journal = {Phys. Rev. B},
  volume = {68},
  issue = {12},
  pages = {125307},
  numpages = {12},
  year = {2003},
  month = {Sep},
  publisher = {American Physical Society},
  doi = {10.1103/PhysRevB.68.125307},
  url = {https://link.aps.org/doi/10.1103/PhysRevB.68.125307}
}

@article{Wan_PRB_2008,
  title = {Fractional quantum Hall effect at $\nu=5/2$: Ground states, non-Abelian quasiholes, and edge modes in a microscopic model},
  author = {Wan, Xin and Hu, Zi-Xiang and Rezayi, E. H. and Yang, Kun},
  journal = {Phys. Rev. B},
  volume = {77},
  issue = {16},
  pages = {165316},
  numpages = {15},
  year = {2008},
  month = {Apr},
  publisher = {American Physical Society},
  doi = {10.1103/PhysRevB.77.165316},
  url = {https://link.aps.org/doi/10.1103/PhysRevB.77.165316}
}

@article{MetropolisTeller_JChemPhys_1953,
author = {Metropolis,Nicholas  and Rosenbluth,Arianna W.  and Rosenbluth,Marshall N.  and Teller,Augusta H.  and Teller,Edward },
title = {Equation of State Calculations by Fast Computing Machines},
journal = {The Journal of Chemical Physics},
volume = {21},
number = {6},
pages = {1087-1092},
year = {1953},
doi = {10.1063/1.1699114},
URL = { https://doi.org/10.1063/1.1699114},
eprint = { https://doi.org/10.1063/1.1699114}
}

@article{MetropolisUlam_JStat_1949,
author = { Nicholas   Metropolis  and  S.   Ulam },
title = {The Monte Carlo Method},
journal = {Journal of the American Statistical Association},
volume = {44},
number = {247},
pages = {335-341},
year  = {1949},
publisher = {Taylor & Francis},
doi = {10.1080/01621459.1949.10483310},
note ={PMID: 18139350}
}

@article{Hastings_Biometrika_1970,
 ISSN = {00063444},
 URL = {http://www.jstor.org/stable/2334940},
 author = {W. K. Hastings},
 journal = {Biometrika},
 number = {1},
 pages = {97--109},
 publisher = {[Oxford University Press, Biometrika Trust]},
 title = {Monte Carlo Sampling Methods Using Markov Chains and Their Applications},
 urldate = {2023-02-19},
 volume = {57},
 year = {1970}
}

@article{Dubail_PRB_2012,
  title = {Edge-state inner products and real-space entanglement spectrum of trial quantum Hall states},
  author = {Dubail, J. and Read, N. and Rezayi, E. H.},
  journal = {Phys. Rev. B},
  volume = {86},
  issue = {24},
  pages = {245310},
  numpages = {32},
  year = {2012},
  month = {Dec},
  publisher = {American Physical Society},
  doi = {10.1103/PhysRevB.86.245310},
  url = {https://link.aps.org/doi/10.1103/PhysRevB.86.245310}
}

@article{BernevigHaldane_PRL_2008,
  title = {Model Fractional Quantum Hall States and Jack Polynomials},
  author = {Bernevig, B. Andrei and Haldane, F. D. M.},
  journal = {Phys. Rev. Lett.},
  volume = {100},
  issue = {24},
  pages = {246802},
  numpages = {4},
  year = {2008},
  month = {Jun},
  publisher = {American Physical Society},
  doi = {10.1103/PhysRevLett.100.246802},
  url = {https://link.aps.org/doi/10.1103/PhysRevLett.100.246802}
}

@article{PriceLamacraft_PRB_2014,
  title = {Fine structure of the phonon in one dimension from quantum hydrodynamics},
  author = {Price, Tom and Lamacraft, Austen},
  journal = {Phys. Rev. B},
  volume = {90},
  issue = {24},
  pages = {241415},
  numpages = {5},
  year = {2014},
  month = {Dec},
  publisher = {American Physical Society},
  doi = {10.1103/PhysRevB.90.241415},
  url = {https://link.aps.org/doi/10.1103/PhysRevB.90.241415}
}

@article{Haldane_PRL_1991,
  title = {``Fractional statistics'' in arbitrary dimensions: A generalization of the Pauli principle},
  author = {Haldane, F. D. M.},
  journal = {Phys. Rev. Lett.},
  volume = {67},
  issue = {8},
  pages = {937--940},
  numpages = {0},
  year = {1991},
  month = {Aug},
  publisher = {American Physical Society},
  doi = {10.1103/PhysRevLett.67.937},
  url = {https://link.aps.org/doi/10.1103/PhysRevLett.67.937}
}

@misc{PriceLamacraft_arxiv_2015,
  doi = {10.48550/ARXIV.1509.08332},
  url = {https://arxiv.org/abs/1509.08332},
  author = {Price, Tom and Lamacraft, Austen},
  title = {Quantum Hydrodynamics in One Dimension beyond the Luttinger Liquid},
  publisher = {arXiv},
  year = {2015}
}

@article{WilkinGunn_PRL_1998,
  title = {Do Attractive Bosons Condense?},
  author = {Wilkin, N. K. and Gunn, J. M. F. and Smith, R. A.},
  journal = {Phys. Rev. Lett.},
  volume = {80},
  issue = {11},
  pages = {2265--2268},
  numpages = {0},
  year = {1998},
  month = {Mar},
  publisher = {American Physical Society},
  doi = {10.1103/PhysRevLett.80.2265},
  url = {https://link.aps.org/doi/10.1103/PhysRevLett.80.2265}
}

@article{Haldane_PRL_1983,
  title = {Fractional Quantization of the Hall Effect: A Hierarchy of Incompressible Quantum Fluid States},
  author = {Haldane, F. D. M.},
  journal = {Phys. Rev. Lett.},
  volume = {51},
  issue = {7},
  pages = {605--608},
  numpages = {0},
  year = {1983},
  month = {Aug},
  publisher = {American Physical Society},
  doi = {10.1103/PhysRevLett.51.605},
  url = {https://link.aps.org/doi/10.1103/PhysRevLett.51.605}
}

@article{TrugmanKivelson_PRB_1985,
  title = {Exact results for the fractional quantum Hall effect with general interactions},
  author = {Trugman, S. A. and Kivelson, S.},
  journal = {Phys. Rev. B},
  volume = {31},
  issue = {8},
  pages = {5280--5284},
  numpages = {0},
  year = {1985},
  month = {Apr},
  publisher = {American Physical Society},
  doi = {10.1103/PhysRevB.31.5280},
  url = {https://link.aps.org/doi/10.1103/PhysRevB.31.5280}
}

@article{SimonRezayiCooper_PRB_2007_2,
  title = {Pseudopotentials for multiparticle interactions in the quantum Hall regime},
  author = {Simon, Steven H. and Rezayi, E. H. and Cooper, Nigel R.},
  journal = {Phys. Rev. B},
  volume = {75},
  issue = {19},
  pages = {195306},
  numpages = {11},
  year = {2007},
  month = {May},
  publisher = {American Physical Society},
  doi = {10.1103/PhysRevB.75.195306},
  url = {https://link.aps.org/doi/10.1103/PhysRevB.75.195306}
}

@article{Laughlin_PRL_1983,
  title = {Anomalous Quantum Hall Effect: An Incompressible Quantum Fluid with Fractionally Charged Excitations},
  author = {Laughlin, R. B.},
  journal = {Phys. Rev. Lett.},
  volume = {50},
  issue = {18},
  pages = {1395--1398},
  numpages = {0},
  year = {1983},
  month = {May},
  publisher = {American Physical Society},
  doi = {10.1103/PhysRevLett.50.1395},
  url = {https://link.aps.org/doi/10.1103/PhysRevLett.50.1395}
}

@article{LeeHuWan_PRB_2014,
  title = {Construction of edge states in fractional quantum Hall systems by Jack polynomials},
  author = {Lee, Ki Hoon and Hu, Zi-Xiang and Wan, Xin},
  journal = {Phys. Rev. B},
  volume = {89},
  issue = {16},
  pages = {165124},
  numpages = {9},
  year = {2014},
  month = {Apr},
  publisher = {American Physical Society},
  doi = {10.1103/PhysRevB.89.165124},
  url = {https://link.aps.org/doi/10.1103/PhysRevB.89.165124}
}

@article{Stone_PRB_1990,
  title = {Schur functions, chiral bosons, and the quantum-Hall-effect edge states},
  author = {Stone, Michael},
  journal = {Phys. Rev. B},
  volume = {42},
  issue = {13},
  pages = {8399--8404},
  numpages = {0},
  year = {1990},
  month = {Nov},
  publisher = {American Physical Society},
  doi = {10.1103/PhysRevB.42.8399},
  url = {https://link.aps.org/doi/10.1103/PhysRevB.42.8399}
}

@article{Avron_PRL_1995,
  title = {Viscosity of Quantum Hall Fluids},
  author = {Avron, J. E. and Seiler, R. and Zograf, P. G.},
  journal = {Phys. Rev. Lett.},
  volume = {75},
  issue = {4},
  pages = {697--700},
  numpages = {0},
  year = {1995},
  month = {Jul},
  publisher = {American Physical Society},
  doi = {10.1103/PhysRevLett.75.697},
  url = {https://link.aps.org/doi/10.1103/PhysRevLett.75.697}
}

@article{Fern_PRB_2018,
  title = {Effective edge state dynamics in the fractional quantum Hall effect},
  author = {Fern, R. and Bondesan, R. and Simon, S. H.},
  journal = {Phys. Rev. B},
  volume = {98},
  issue = {15},
  pages = {155321},
  numpages = {20},
  year = {2018},
  month = {Oct},
  publisher = {American Physical Society},
  doi = {10.1103/PhysRevB.98.155321},
  url = {https://link.aps.org/doi/10.1103/PhysRevB.98.155321}
}

@article{ImambekovGlazman_Science_2009,
	doi = {10.1126/science.1165403},
	url = {https://doi.org/10.1126%2Fscience.1165403},
	year = 2009,
	month = {jan},
	publisher = {American Association for the Advancement of Science ({AAAS})},
	volume = {323},
	number = {5911},
	pages = {228--231},
	author = {Adilet Imambekov and Leonid I. Glazman},
	title = {Universal Theory of Nonlinear Luttinger Liquids}, 
	journal = {Science}
}

@book{jainCompositeFermionsBook_2007, 
place={Cambridge}, 
title={Composite Fermions}, 
DOI={10.1017/CBO9780511607561}, 
publisher={Cambridge University Press}, 
author={Jain, Jainendra K.}, 
year={2007}}

@article{Tsui_PRL_1982,
  title = {Two-Dimensional Magnetotransport in the Extreme Quantum Limit},
  author = {Tsui, D. C. and Stormer, H. L. and Gossard, A. C.},
  journal = {Phys. Rev. Lett.},
  volume = {48},
  issue = {22},
  pages = {1559--1562},
  numpages = {0},
  year = {1982},
  month = {May},
  publisher = {American Physical Society},
  doi = {10.1103/PhysRevLett.48.1559},
  url = {https://link.aps.org/doi/10.1103/PhysRevLett.48.1559}
}

@article{Bartolomei_science_2020,
       author = {{Bartolomei}, H. and {Kumar}, M. and {Bisognin}, R. and {Marguerite}, A. and {Berroir}, J. -M. and {Bocquillon}, E. and {Pla{\c{c}}ais}, B. and {Cavanna}, A. and {Dong}, Q. and {Gennser}, U. and {Jin}, Y. and {F{\`e}ve}, G.},
        title = "{Fractional statistics in anyon collisions}",
      journal = {Science},
         year = 2020,
        month = apr,
       volume = {368},
       number = {6487},
        pages = {173-177},
          doi = {10.1126/science.aaz5601},
archivePrefix = {arXiv},
       eprint = {2006.13157},
 primaryClass = {cond-mat.mes-hall}
}

@article{Wen_AdvPhys_1995,
    author = "Wen, Xiao-Gang",
    title = "{Topological orders and edge excitations in fractional quantum Hall states}",
    eprint = "cond-mat/9506066",
    archivePrefix = "arXiv",
    reportNumber = "PRINT-95-148 (MIT)",
    doi = "10.1080/00018739500101566",
    journal = "Adv. Phys.",
    volume = "44",
    number = "5",
    pages = "405--473",
    year = "1995"
}

@book{prangeQHE2012,
  title={The Quantum Hall Effect},
  author={Prange, R.E. and Cage, M.E. and Klitzing, K. and Girvin, S.M. and Chang, A.M. and Duncan, F. and Haldane, M. and Laughlin, R.B. and Pruisken, A.M.M. and Thouless, D.J.},
  isbn={9781461233503},
  series={Graduate Texts in Contemporary Physics},
  url={https://books.google.it/books?id=mxrSBwAAQBAJ},
  year={2012},
  publisher={Springer New York}
}

@article{Nayak_RMP_2008,
	doi = {10.1103/revmodphys.80.1083},
	url = {https://doi.org/10.1103%2Frevmodphys.80.1083},
	year = 2008,
	month = {sep},
	publisher = {American Physical Society ({APS})},
	volume = {80},
	number = {3},
	pages = {1083--1159},
	author = {Chetan Nayak and Steven H. Simon and Ady Stern and Michael Freedman and Sankar Das Sarma},
	title = {Non-Abelian anyons and topological quantum computation},
	journal = {Reviews of Modern Physics}
}

@article{Jolad_PRB_2007,
  title = {"Electron operator at the edge of the $1 /3$ fractional quantum Hall liquid"},
  author = {Jolad, Shivakumar and Chang, Chia-Chen and Jain, Jainendra K.},
  journal = {Phys. Rev. B},
  volume = {75},
  issue = {16},
  pages = {165306},
  numpages = {8},
  year = {2007},
  month = {Apr},
  publisher = {American Physical Society},
  doi = {10.1103/PhysRevB.75.165306},
  url = {https://link.aps.org/doi/10.1103/PhysRevB.75.165306}
}

@article{Jolad_PRL_2009,
  title = {Testing the Topological Nature of the Fractional Quantum Hall Edge},
  author = {Jolad, Shivakumar and Jain, Jainendra K.},
  journal = {Phys. Rev. Lett.},
  volume = {102},
  issue = {11},
  pages = {116801},
  numpages = {4},
  year = {2009},
  month = {Mar},
  publisher = {American Physical Society},
  doi = {10.1103/PhysRevLett.102.116801},
  url = {https://link.aps.org/doi/10.1103/PhysRevLett.102.116801}
}

@article{Jolad_PRB_2010,
  title = {Fractional quantum Hall edge: Effect of nonlinear dispersion and edge roton},
  author = {Jolad, Shivakumar and Sen, Diptiman and Jain, Jainendra K.},
  journal = {Phys. Rev. B},
  volume = {82},
  issue = {7},
  pages = {075315},
  numpages = {14},
  year = {2010},
  month = {Aug},
  publisher = {American Physical Society},
  doi = {10.1103/PhysRevB.82.075315},
  url = {https://link.aps.org/doi/10.1103/PhysRevB.82.075315}
}

@article{Wen_PRL_1990,
  title = {Electrodynamical properties of gapless edge excitations in the fractional quantum Hall states},
  author = {Wen, X. G.},
  journal = {Phys. Rev. Lett.},
  volume = {64},
  issue = {18},
  pages = {2206--2209},
  numpages = {0},
  year = {1990},
  month = {Apr},
  publisher = {American Physical Society},
  doi = {10.1103/PhysRevLett.64.2206},
  url = {https://link.aps.org/doi/10.1103/PhysRevLett.64.2206}
}

@article{Wen_PRB_1990b,
  title = {Chiral Luttinger liquid and the edge excitations in the fractional quantum Hall states},
  author = {Wen, X. G.},
  journal = {Phys. Rev. B},
  volume = {41},
  issue = {18},
  pages = {12838--12844},
  numpages = {0},
  year = {1990},
  month = {Jun},
  publisher = {American Physical Society},
  doi = {10.1103/PhysRevB.41.12838},
  url = {https://link.aps.org/doi/10.1103/PhysRevB.41.12838}
}

@article{Wen_PRB_1991b,
  title = {Edge transport properties of the fractional quantum Hall states and weak-impurity scattering of a one-dimensional charge-density wave},
  author = {Wen, Xiao-Gang},
  journal = {Phys. Rev. B},
  volume = {44},
  issue = {11},
  pages = {5708--5719},
  numpages = {0},
  year = {1991},
  month = {Sep},
  publisher = {American Physical Society},
  doi = {10.1103/PhysRevB.44.5708},
  url = {https://link.aps.org/doi/10.1103/PhysRevB.44.5708}
}

@article{Wen_PRB_1991,
  title = {Gapless boundary excitations in the quantum Hall states and in the chiral spin states},
  author = {Wen, X. G.},
  journal = {Phys. Rev. B},
  volume = {43},
  issue = {13},
  pages = {11025--11036},
  numpages = {0},
  year = {1991},
  month = {May},
  publisher = {American Physical Society},
  doi = {10.1103/PhysRevB.43.11025},
  url = {https://link.aps.org/doi/10.1103/PhysRevB.43.11025}
}

@article{Cooper_2008,
	doi = {10.1080/00018730802564122},
	url = {https://doi.org/10.1080%2F00018730802564122},
	year = 2008,
	month = {nov},
	publisher = {Informa {UK} Limited},
	volume = {57},
	number = {6},
	pages = {539--616},
	author = {N.R. Cooper},
	title = {Rapidly rotating atomic gases},
	journal = {Advances in Physics}
}

@article{Bloch_RMP_2008,
  title = {Many-body physics with ultracold gases},
  author = {Bloch, Immanuel and Dalibard, Jean and Zwerger, Wilhelm},
  journal = {Rev. Mod. Phys.},
  volume = {80},
  issue = {3},
  pages = {885--964},
  numpages = {0},
  year = {2008},
  month = {Jul},
  publisher = {American Physical Society},
  doi = {10.1103/RevModPhys.80.885},
  url = {https://link.aps.org/doi/10.1103/RevModPhys.80.885}
}

@article{Goldman_RepProgPhys_2014,
	doi = {10.1088/0034-4885/77/12/126401},
	url = {https://doi.org/10.1088%2F0034-4885%2F77%2F12%2F126401},
	year = 2014,
	month = {nov},
	publisher = {{IOP} Publishing},
	volume = {77},
	number = {12},
	pages = {126401},
	author = {N Goldman and G Juzeli{\={u}
}nas and P Öhberg and I B Spielman},
	title = {Light-induced gauge fields for ultracold atoms}, 
	journal = {Reports on Progress in Physics}
}

@article{Carusotto_RMP_2013,
  title = {Quantum fluids of light},
  author = {Carusotto, Iacopo and Ciuti, Cristiano},
  journal = {Rev. Mod. Phys.},
  volume = {85},
  issue = {1},
  pages = {299--366},
  numpages = {0},
  year = {2013},
  month = {Feb},
  publisher = {American Physical Society},
  doi = {10.1103/RevModPhys.85.299},
  url = {https://link.aps.org/doi/10.1103/RevModPhys.85.299}
}

@article{Ozawa_RMP_2019,
  title = {Topological photonics},
  author = {Ozawa, Tomoki and Price, Hannah M. and Amo, Alberto and Goldman, Nathan and Hafezi, Mohammad and Lu, Ling and Rechtsman, Mikael C. and Schuster, David and Simon, Jonathan and Zilberberg, Oded and Carusotto, Iacopo},
  journal = {Rev. Mod. Phys.},
  volume = {91},
  issue = {1},
  pages = {015006},
  numpages = {76},
  year = {2019},
  month = {Mar},
  publisher = {American Physical Society},
  doi = {10.1103/RevModPhys.91.015006},
  url = {https://link.aps.org/doi/10.1103/RevModPhys.91.015006}
}

@article{Carusotto_NatPhys_2020,
title = "Photonic materials in circuit quantum electrodynamics",
author = "Iacopo Carusotto and Houck, {Andrew A.} and Koll{\'a}r, {Alicia J.} and Pedram Roushan and Schuster, {David I.} and Jonathan Simon",
year = "2020",
month = mar,
day = "1",
doi = "10.1038/s41567-020-0815-y",
volume = "16",
pages = "268--279",
journal = "Nature Physics",
issn = "1745-2473",
publisher = "Nature Publishing Group",
number = "3",
}

@article{Schine_Nat_2016,
       author = {{Schine}, Nathan and {Ryou}, Albert and {Gromov}, Andrey and {Sommer}, Ariel and {Simon}, Jonathan},
        title = "{Synthetic Landau levels for photons}",
      journal = {\nat},
         year = 2016,
        month = jun,
       volume = {534},
       number = {7609},
        pages = {671-675},
          doi = {10.1038/nature17943},
archivePrefix = {arXiv},
       eprint = {1511.07381},
 primaryClass = {cond-mat.quant-gas},
       adsurl = {https://ui.adsabs.harvard.edu/abs/2016Natur.534..671S}
}

@article{Ashoori_PRB_1992,
       author = {{Ashoori}, R.~C. and {Stormer}, H.~L. and {Pfeiffer}, L.~N. and {Baldwin}, K.~W. and {West}, K.},
        title = "{Edge magnetoplasmons in the time domain}",
      journal = {\prb},
         year = 1992,
        month = feb,
       volume = {45},
       number = {7},
        pages = {3894-3897},
          doi = {10.1103/PhysRevB.45.3894},
       adsurl = {https://ui.adsabs.harvard.edu/abs/1992PhRvB..45.3894A}
}

@article{Girvin_PRB_1986,
  title = {Magneto-roton theory of collective excitations in the fractional quantum Hall effect},
  author = {Girvin, S. M. and MacDonald, A. H. and Platzman, P. M.},
  journal = {Phys. Rev. B},
  volume = {33},
  issue = {4},
  pages = {2481--2494},
  numpages = {0},
  year = {1986},
  month = {Feb},
  publisher = {American Physical Society},
  doi = {10.1103/PhysRevB.33.2481},
  url = {https://link.aps.org/doi/10.1103/PhysRevB.33.2481}
}

@article{Khodas_PRB_2007,
  title = {Fermi-Luttinger liquid: Spectral function of interacting one-dimensional fermions},
  author = {Khodas, M. and Pustilnik, M. and Kamenev, A. and Glazman, L. I.},
  journal = {Phys. Rev. B},
  volume = {76},
  issue = {15},
  pages = {155402},
  numpages = {21},
  year = {2007},
  month = {Oct},
  publisher = {American Physical Society},
  doi = {10.1103/PhysRevB.76.155402},
  url = {https://link.aps.org/doi/10.1103/PhysRevB.76.155402}
}

@ARTICLE{MooreRead_NPB_1991,
       author = {{Moore}, Gregory and {Read}, Nicholas},
        title = "{Nonabelions in the fractional quantum hall effect}",
      journal = {Nuclear Physics B},
         year = 1991,
        month = aug,
       volume = {360},
       number = {2},
        pages = {362-396},
          doi = {10.1016/0550-3213(91)90407-O},
       adsurl = {https://ui.adsabs.harvard.edu/abs/1991NuPhB.360..362M}
}

@article{Yusa_PRR_2022,
       author = {{Kamiyama}, Akinori and {Matsuura}, Masahiro and {Moore}, John N. and {Mano}, Takaaki and {Shibata}, Naokazu and {Yusa}, Go},
        title = "{Real-time and space visualization of excitations of the {\ensuremath{\nu}} =1 /3 fractional quantum Hall edge}",
      journal = {Physical Review Research},
         year = 2022,
        month = mar,
       volume = {4},
       number = {1},
          eid = {L012040},
        pages = {L012040},
          doi = {10.1103/PhysRevResearch.4.L012040},
archivePrefix = {arXiv},
       eprint = {2201.10180},
 primaryClass = {cond-mat.mes-hall},
       adsurl = {https://ui.adsabs.harvard.edu/abs/2022PhRvR...4a2040K}
}

@article{Umansky_NatPhys_2023,
       author = {{Aharon Melcer}, Ron and {Konyzheva}, Sofia and {Heiblum}, Moty and {Umansky}, Vladimir},
        title = "{Direct Determination of the Topological Thermal Conductance via Local Power Measurement}",
        journal = {Nature Physics},
        year = {2023},
        journal = "Nature Physics",
        issn = "1745-2481",
        doi = {https://doi.org/10.1038/s41567-022-01885-5}
}

@article{Fletcher_Science_2021,
	doi = {10.1126/science.aba7202},
	url = {https://doi.org/10.1126%2Fscience.aba7202},
	year = 2021,
	month = {jun},
	publisher = {American Association for the Advancement of Science ({AAAS})},
	volume = {372},
	number = {6548},
	pages = {1318--1322},
	author = {Richard J. Fletcher and Airlia Shaffer and Cedric C. Wilson and Parth B. Patel and Zhenjie Yan and Valentin Cr{\'{e}
}pel and Biswaroop Mukherjee and Martin W. Zwierlein},
	title = {Geometric squeezing into the lowest Landau level},
	journal = {Science}
}

@article{Clark_Nat_2020,
       author = {{Clark}, Logan W. and {Schine}, Nathan and {Baum}, Claire and {Jia}, Ningyuan and {Simon}, Jonathan},
        title = "{Observation of Laughlin states made of light}",
      journal = {\nat},
         year = 2020,
        month = jun,
       volume = {582},
       number = {7810},
        pages = {41-45},
          doi = {10.1038/s41586-020-2318-5},
archivePrefix = {arXiv},
       eprint = {1907.05872},
 primaryClass = {cond-mat.quant-gas},
       adsurl = {https://ui.adsabs.harvard.edu/abs/2020Natur.582...41C}
}

@article{Roushan_NatPhys_2017,
       author = {{Roushan}, P. and {Neill}, C. and {Megrant}, A. and {Chen}, Y. and {Babbush}, R. and {Barends}, R. and {Campbell}, B. and {Chen}, Z. and {Chiaro}, B. and {Dunsworth}, A. and {Fowler}, A. and {Jeffrey}, E. and {Kelly}, J. and {Lucero}, E. and {Mutus}, J. and {O'Malley}, P.~J.~J. and {Neeley}, M. and {Quintana}, C. and {Sank}, D. and {Vainsencher}, A. and {Wenner}, J. and {White}, T. and {Kapit}, E. and {Neven}, H. and {Martinis}, J.},
        title = "{Chiral ground-state currents of interacting photons in a synthetic magnetic field}",
      journal = {Nature Physics},
         year = 2017,
        month = feb,
       volume = {13},
       number = {2},
        pages = {146-151},
          doi = {10.1038/nphys3930},
archivePrefix = {arXiv},
       eprint = {1606.00077},
 primaryClass = {quant-ph},
       adsurl = {https://ui.adsabs.harvard.edu/abs/2017NatPh..13..146R}
}

@article{Fabbri_PRA_2015,
  title = {Dynamical structure factor of one-dimensional Bose gases: Experimental signatures of beyond-Luttinger-liquid physics},
  author = {Fabbri, N. and Panfil, M. and Cl\'ement, D. and Fallani, L. and Inguscio, M. and Fort, C. and Caux, J.-S.},
  journal = {Phys. Rev. A},
  volume = {91},
  issue = {4},
  pages = {043617},
  numpages = {7},
  year = {2015},
  month = {Apr},
  publisher = {American Physical Society},
  doi = {10.1103/PhysRevA.91.043617},
  url = {https://link.aps.org/doi/10.1103/PhysRevA.91.043617}
}

@article{Leonard_2022,
   title={Realization of a fractional quantum Hall state with ultracold atoms},
   volume={619},
   ISSN={1476-4687},
   url={http://dx.doi.org/10.1038/s41586-023-06122-4},
   DOI={10.1038/s41586-023-06122-4},
   number={7970},
   journal={Nature},
   publisher={Springer Science and Business Media LLC},
   author={Léonard, Julian and Kim, Sooshin and Kwan, Joyce and Segura, Perrin and Grusdt, Fabian and Repellin, Cécile and Goldman, Nathan and Greiner, Markus},
   year={2023},
   month=jun, pages={495–499} }

@article{Stone_PRB_1992,
  title = {Edge waves in the quantum Hall effect and quantum dots},
  author = {Stone, Michael and Wyld, H. W. and Schult, R. L.},
  journal = {Phys. Rev. B},
  volume = {45},
  issue = {24},
  pages = {14156--14161},
  numpages = {0},
  year = {1992},
  month = {Jun},
  publisher = {American Physical Society},
  doi = {10.1103/PhysRevB.45.14156},
  url = {https://link.aps.org/doi/10.1103/PhysRevB.45.14156}
}

@article{Tai_Nat_2017,
       author = {{Tai}, M. Eric and {Lukin}, Alexander and {Rispoli}, Matthew and {Schittko}, Robert and {Menke}, Tim and {Dan Borgnia} and {Preiss}, Philipp M. and {Grusdt}, Fabian and {Kaufman}, Adam M. and {Greiner}, Markus},
        title = "{Microscopy of the interacting Harper-Hofstadter model in the two-body limit}",
      journal = {\nat},
         year = 2017,
        month = jun,
       volume = {546},
       number = {7659},
        pages = {519-523},
          doi = {10.1038/nature22811},
archivePrefix = {arXiv},
       eprint = {1612.05631},
 primaryClass = {cond-mat.quant-gas},
       adsurl = {https://ui.adsabs.harvard.edu/abs/2017Natur.546..519T}
}

@misc{Gemelke_2010,
  doi = {10.48550/ARXIV.1007.2677},
  url = {https://arxiv.org/abs/1007.2677},
  author = {Gemelke, Nathan and Sarajlic, Edina and Chu, Steven},
  title = {Rotating Few-body Atomic Systems in the Fractional Quantum Hall Regime},
  publisher = {arXiv},
  year = {2010}, 
  copyright = {arXiv.org perpetual, non-exclusive license}
}

@article{Cooper_RMP_2019,
  title = {Topological bands for ultracold atoms},
  author = {Cooper, N. R. and Dalibard, J. and Spielman, I. B.},
  journal = {Rev. Mod. Phys.},
  volume = {91},
  issue = {1},
  pages = {015005},
  numpages = {55},
  year = {2019},
  month = {Mar},
  publisher = {American Physical Society},
  doi = {10.1103/RevModPhys.91.015005},
  url = {https://link.aps.org/doi/10.1103/RevModPhys.91.015005}
}

@ARTICLE{Banerjee_Nat_2018,
       author = {{Banerjee}, Mitali and {Heiblum}, Moty and {Umansky}, Vladimir and {Feldman}, Dima E. and {Oreg}, Yuval and {Stern}, Ady},
        title = "{Observation of half-integer thermal Hall conductance}",
      journal = {\nat},
         year = 2018,
        month = jun,
       volume = {559},
       number = {7713},
        pages = {205-210},
          doi = {10.1038/s41586-018-0184-1},
archivePrefix = {arXiv},
       eprint = {1710.00492},
 primaryClass = {cond-mat.mes-hall},
       adsurl = {https://ui.adsabs.harvard.edu/abs/2018Natur.559..205B}
}

@article{Wen_science_2019,
author = {Xiao-Gang Wen },
title = {Choreographed entanglement dances: Topological states of quantum matter},
journal = {Science},
volume = {363},
number = {6429},
pages = {eaal3099},
year = {2019},
doi = {10.1126/science.aal3099},
URL = {https://www.science.org/doi/abs/10.1126/science.aal3099},
eprint = {https://www.science.org/doi/pdf/10.1126/science.aal3099}}

@article{dePicciotto_nat_1997,
       author = {{de-Picciotto}, R. and {Reznikov}, M. and {Heiblum}, M. and {Umansky}, V. and {Bunin}, G. and {Mahalu}, D.},
        title = "{Direct observation of a fractional charge}",
      journal = {\nat},
         year = 1997,
        month = sep,
       volume = {389},
       number = {6647},
        pages = {162-164},
          doi = {10.1038/38241}
}

@article{Arovas_PRL_1984,
  title = {Fractional Statistics and the Quantum Hall Effect},
  author = {Arovas, Daniel and Schrieffer, J. R. and Wilczek, Frank},
  journal = {Phys. Rev. Lett.},
  volume = {53},
  issue = {7},
  pages = {722--723},
  numpages = {0},
  year = {1984},
  month = {Aug},
  publisher = {American Physical Society},
  doi = {10.1103/PhysRevLett.53.722},
  url = {https://link.aps.org/doi/10.1103/PhysRevLett.53.722}
}

@ARTICLE{Nakamura_Nature_2020,
       author = {{Nakamura}, J. and {Liang}, S. and {Gardner}, G.~C. and {Manfra}, M.~J.},
        title = "{Direct observation of anyonic braiding statistics}",
      journal = {Nature Physics},
         year = 2020,
        month = sep,
       volume = {16},
       number = {9},
        pages = {931-936},
          doi = {10.1038/s41567-020-1019-1},
archivePrefix = {arXiv},
       eprint = {2006.14115}
}

@article{Bettelheim_PRL_2006,
  title = {Nonlinear Quantum Shock Waves in Fractional Quantum Hall Edge States},
  author = {Bettelheim, E. and Abanov, Alexander G. and Wiegmann, P.},
  journal = {Phys. Rev. Lett.},
  volume = {97},
  issue = {24},
  pages = {246401},
  numpages = {4},
  year = {2006},
  month = {Dec},
  publisher = {American Physical Society},
  doi = {10.1103/PhysRevLett.97.246401},
  url = {https://link.aps.org/doi/10.1103/PhysRevLett.97.246401}
}

@article{Wiegmann_PRL_2012,
  title = {Nonlinear Hydrodynamics and Fractionally Quantized Solitons at the Fractional Quantum Hall Edge},
  author = {Wiegmann, P.},
  journal = {Phys. Rev. Lett.},
  volume = {108},
  issue = {20},
  pages = {206810},
  numpages = {5},
  year = {2012},
  month = {May},
  publisher = {American Physical Society},
  doi = {10.1103/PhysRevLett.108.206810},
  url = {https://link.aps.org/doi/10.1103/PhysRevLett.108.206810}
}

@book{SimonWavefunctionology_2020,
	doi = {10.1142/11751},
	url = {https://doi.org/10.1142%2F11751},
	year = 2020,
	month = {jan},
	publisher = {{WORLD} {SCIENTIFIC}},
	author = {Bertrand I Halperin and Jainendra K Jain},
	title = {Fractional Quantum Hall Effects}
}

@article{Chang_RMP_2003,
  title = {Chiral Luttinger liquids at the fractional quantum Hall edge},
  author = {Chang, A. M.},
  journal = {Rev. Mod. Phys.},
  volume = {75},
  issue = {4},
  pages = {1449--1505},
  numpages = {0},
  year = {2003},
  month = {Nov},
  publisher = {American Physical Society},
  doi = {10.1103/RevModPhys.75.1449},
  url = {https://link.aps.org/doi/10.1103/RevModPhys.75.1449}
}

@article{Cazalilla_PRA_2003,
  title = {Surface modes of ultracold atomic clouds with a very large number of vortices},
  author = {Cazalilla, M. A.},
  journal = {Phys. Rev. A},
  volume = {67},
  issue = {6},
  pages = {063613},
  numpages = {22},
  year = {2003},
  month = {Jun},
  publisher = {American Physical Society},
  doi = {10.1103/PhysRevA.67.063613},
  url = {https://link.aps.org/doi/10.1103/PhysRevA.67.063613}
}

@article{SimonRezayiCooper_PRB_2007_1,
  title = {Generalized quantum Hall projection Hamiltonians},
  author = {Simon, Steven H. and Rezayi, E. H. and Cooper, Nigel R.},
  journal = {Phys. Rev. B},
  volume = {75},
  issue = {7},
  pages = {075318},
  numpages = {8},
  year = {2007},
  month = {Feb},
  publisher = {American Physical Society},
  doi = {10.1103/PhysRevB.75.075318},
  url = {https://link.aps.org/doi/10.1103/PhysRevB.75.075318}
}

@article{KunYang_PRL_2003,
  title = {Field Theoretical Description of Quantum Hall Edge Reconstruction},
  author = {Yang, Kun},
  journal = {Phys. Rev. Lett.},
  volume = {91},
  issue = {3},
  pages = {036802},
  numpages = {4},
  year = {2003},
  month = {Jul},
  publisher = {American Physical Society},
  doi = {10.1103/PhysRevLett.91.036802},
  url = {https://link.aps.org/doi/10.1103/PhysRevLett.91.036802}
}

@article{Umucalilar_PRA_2017,
  title = {Generation and spectroscopic signatures of a fractional quantum Hall liquid of photons in an incoherently pumped optical cavity},
  author = {Umucal\ifmmode \imath \else \i \fi{}lar, R. O. and Carusotto, I.},
  journal = {Phys. Rev. A},
  volume = {96},
  issue = {5},
  pages = {053808},
  numpages = {11},
  year = {2017},
  month = {Nov},
  publisher = {American Physical Society},
  doi = {10.1103/PhysRevA.96.053808},
  url = {https://link.aps.org/doi/10.1103/PhysRevA.96.053808}
}

@article{Nardin_PRA_2023,
  title = {Linear and nonlinear edge dynamics of trapped fractional quantum Hall droplets},
  author = {Nardin, Alberto and Carusotto, Iacopo},
  journal = {Phys. Rev. A},
  volume = {107},
  issue = {3},
  pages = {033320},
  numpages = {13},
  year = {2023},
  month = {Mar},
  publisher = {American Physical Society},
  doi = {10.1103/PhysRevA.107.033320},
  url = {https://link.aps.org/doi/10.1103/PhysRevA.107.033320}
}

@article{Nardin_EPL_2020,
doi = {10.1209/0295-5075/132/10002},
url = {https://dx.doi.org/10.1209/0295-5075/132/10002},
year = {2020},
month = {nov},
publisher = {EDP Sciences, IOP Publishing and Società Italiana di Fisica},
volume = {132},
number = {1},
pages = {10002},
author = {Alberto Nardin and Iacopo Carusotto},
title = {Non-linear edge dynamics of an integer quantum Hall fluid},
journal = {Europhysics Letters}
}

@article{Gross2021,
  title={Quantum gas microscopy for single atom and spin detection},
  author={Gross, Christian and Bakr, Waseem S},
  journal={Nature Physics},
  volume={17},
  number={12},
  pages={1316--1323},
  year={2021},
  publisher={Nature Publishing Group UK London}
}

@article{Gauthier2016,
  title={Direct imaging of a digital-micromirror device for configurable microscopic optical potentials},
  author={Gauthier, G and Lenton, I and Parry, N McKay and Baker, M and Davis, MJ and Rubinsztein-Dunlop, H and Neely, TW},
  journal={Optica},
  volume={3},
  number={10},
  pages={1136--1143},
  year={2016},
  publisher={Optica Publishing Group}
}

@article{Frigerio_CommPhys_2024,
  title={Gate tunable edge magnetoplasmon resonators},
  author={Frigerio, Elric and Rebora, Giacomo and Ruelle, M{\'e}lanie and Souquet-Basi{\`e}ge, Hubert and Jin, Yong and Gennser, Ulf and Cavanna, Antonella and Pla{\c{c}}ais, Bernard and Baudin, Emmanuel and Berroir, Jean-Marc and others},
  journal={Communications Physics},
  volume={7},
  number={1},
  pages={314},
  year={2024},
  publisher={Nature Publishing Group UK London}
}
    
\end{document}